\documentclass[trackchanges]{aastex701}
	\shorttitle{Shocks, compressible perturbations, and intermittency in the very local interstellar medium}
	\shortauthors{Fraternale et al.}
	\accepted{\today}
	\submitjournal{ApJ}
	\graphicspath{{./}{figures/}}

		\usepackage{amsmath}
		\usepackage{amsbsy}
		\usepackage{amsfonts}
		\usepackage{amssymb}
		\usepackage{bm}
		\usepackage{placeins}
		\usepackage{xcolor}
		\definecolor{orangered}{HTML}{FF7110}
		\definecolor{green}{HTML}{00CC00}
		\usepackage{xspace}

		\newcommand{\au}{\ensuremath{\mathrm{au}}\xspace}
		\newcommand{\lism}{\ensuremath{\mathrm{LISM}}\xspace}
		\newcommand{\He}{\ensuremath{\mathrm{He}}\xspace}
		\newcommand{\HeI}{\ensuremath{{\mathrm{He}^+}}\xspace}
		
		\newcommand{\alp}{\ensuremath{{\mathrm{He}^{2+}}}\xspace}
		\newcommand{\HH}{\ensuremath{\mathrm{H}}\xspace}
		
		\newcommand{\p}{\ensuremath{\mathrm{p}}\xspace}
		\newcommand{\e}{\ensuremath{\mathrm{e}}\xspace}
		\newcommand{\magn}{\ensuremath{\mathrm{mag}}\xspace}
		
		\newcommand{\dyn}{\ensuremath{\mathrm{dyn}}\xspace}
		\newcommand{\tth}{\ensuremath{\mathrm{th}}\xspace}
		\newcommand{\pui}{\ensuremath{\mathrm{PUI}}\xspace}
		\newcommand{\puis}{\ensuremath{\mathrm{PUIs}}\xspace}
		\newcommand{\pth}{\ensuremath{\mathrm{p,th}}\xspace}

		\newcommand{\cc}{\ensuremath{\mathrm{cm^{-3}}}\xspace}

\begin{document}

				\title{Shocks, compressible perturbations, and intermittency in the very local interstellar medium: Voyager 1 and 2 observations and 
					numerical modeling}

				\author[0000-0002-4700-2762]{Federico Fraternale}
				\affiliation{Center for Space Plasma and Aeronomic Research, University of Alabama in Huntsville, Huntsville, AL 35899, USA}
				\email[show]{federico.fraternale@uah.edu}
				
				\author[0000-0002-6409-2392]{Nikolai V. Pogorelov}
				\affiliation{Department of Space Science, University of Alabama in Huntsville, Huntsville, AL 35805, USA}
				\altaffiliation{Center for Space Plasma and Aeronomic Research, University of Alabama in Huntsville, Huntsville, AL 35899, USA}
				\email{np0002@uah.edu}
				
				\author[0000-0002-4700-2762]{Ratan Bera}
				\affiliation{Department of Physics, School of Physical and Biological Sciences, Manipal University Jaipur, Jaipur, 303007, Rajasthan, India}
				\email{ratan.bera@jaipur.manipal.edu}
				
				\author[0000-0002-5569-1553]{Leonard F. Burlaga}
				\affiliation{Leonard F. Burlaga, Inc., Davidsonville, MD 21035, USA}
				\email{lburlagahsp@verizon.net}

				\author[0000-0003-3957-2359]{Maciej Bzowski}
				\affiliation{Space Research Center PAS (CBK PAN), Bartycka 18A, 00-716 Warsaw, Poland}
				\email{bzowski@cbk.waw.pl}

				\begin{abstract}
					
					
					The Voyager spacecraft (V1, V2) provide unique in situ measurements of perturbations propagating beyond the heliopause through the very local interstellar medium (VLISM), including the shocks and pressure fronts whose origin is debated. In particular, a jump in magnetic field strength, observed by V1 in 2020.4 at 149.3 au from the Sun, was followed by a distinct “hump” and persistently strong magnetic field, both requiring theoretical explanation.
					
					This paper offers an interpretation of those observations using a self-consistent, MHD model of the solar wind - LISM interaction driven by the OMNI and interplanetary scintillation data combined with a turbulence analysis of Voyager data.
					
					Our simulations convincingly demonstrate that global, solar-cycle-driven compressions, on hitting the heliopause, can reproduce those puzzling V1 observations. They appear to be associated with solar cycle 24, whereas similar interstellar magnetic field structures can occur once per cycle.
					
					The turbulence analysis reveals time-dependent magnetic compressibility that persists up to 165 au at scales below 10 days. Turbulence intermittency at scales below 1 hour is mostly confined to specific intervals, possibly associated with a broad foreshock region. The apparent disappearance of intermittency since 2022 reflects the turbulence weakening rather than a fundamental change in VLISM properties.
					
					We predict that V1 will record relatively strong magnetic field strengths until $\sim$2030, followed by weaker, infrequent perturbations. At V2, we expect multiple solar-driven compressions before 2026, followed by a major event induced by solar cycle 25 around 2030. New Horizons is expected to cross the termination shock at 80$\pm$2 au in 2031.

				\end{abstract}
				
			\keywords{Heliosphere (711) --- Heliopause (707) --- Solar wind (1534) --- Interstellar medium (847) --- Termination shock (1690) --- Charge exchange ionization (2056) --- Astrosphere interstellar medium interactions (106) --- Heliosheath (710)}
			

			\section{Introduction}\label{sec:intro}
			Voyager 1 and 2 (V1 and V2) were launched in 1977 and have been making observations to the present time. Both spacecraft moved past the heliopause (HP), the boundary between solar and interstellar plasma, which was crossed by V1 on 2012/DOY 238 (August 25) at 121.6 au and by V2 on 2018/DOY 249 (September 5) \citep{stone13, stone2019}. The spacecraft traveled through the heliosphere, across the termination shock (TS), through the hot inner heliosheath (IHS) energetically dominated by pickup ions (PUIs), past the HP, and since then have been moving through the very local interstellar medium \citep[VLISM,][]{zank2015,fraternale2021b}, which we define as the portion of the local interstellar medium (LISM) perturbed by the presence of the heliosphere up to thousands of au from the Sun \citep[e.g., see][]{pogorelov2025}. The spacecraft are now in the interstellar region named the outer heliosheath (OHS), where the interstellar plasma flow is significantly decelerated and heated by a global compression and charge exchange processes, regardless of whether a heliospheric bow shock (BS) exists or not.
			Considering the interstellar scales, the spacecraft are still relatively close to the heliosphere. They are approaching the maxima in the  plasma and neutral atom densities at about 200–250 au from the Sun, according, e.g., to the recent kinetic models of \citet{fraternale2021b, fraternale2024b, fraternale2025a}. Voyager 1 and 2 have now reached heliocentric distances of approximately 172~au and 143~au, respectively.
			In the OHS, Voyager 1 observed two major shocks and two pressure fronts driven by solar activity, particularly by the global merged interaction regions (GMIRs) that impacted the HP on its inner side. Meanwhile, V2 measured two pressure fronts and one shock 
			\citep{burlaga2022,burlaga2023,burlaga2024a,burlaga2024b}. This is shown in Fig.~\ref{fig:data} below.
			
			The second pressure front observed by V1 (\textit{pf2}) in 2020.4  at 149.3~au is particularly interesting and differs from all previous compressions. It was followed by an increase in the interstellar magnetic field (ISMF) strength, B, which in turn was followed by a ``magnetic hump,'' which is located in the time interval between the observed pressure front and 2022.0, with the peak reached in 2021.4 \citep{burlaga2023}. After 2022, V1 observed strong, slowly varying magnetic fields until at least mid-2024 ($\sim$163 au). The azimuthal and elevation angles of B have been declining nearly linearly throughout the OHS to date, although some small deviations from linearity are present after encountering the magnetic hump. 
			
			Global, data-driven, 3D MHD models of the solar wind (SW)–LISM interaction by \citet{kim2017b} were able to reproduce the major features of shocks and their arrival times remarkably well. By processing the \citet{kim2017b} simulation results, \citet{pogorelov2021} showed that a major shock/compression-wave event observed in 2020, was associated with the merger of multiple shocks. This supported the hypothesis about \textit{pf2} being indeed a pressure front of solar origin. This phenomenon was further analyzed by \citet{zirnstein2024}, who used the same simulation results
			with the purpose of finding correlations between some of these shocks and enhancements in energetic neutral atom (ENA) fluxes observed by IBEX.  
			\citet{burlaga2023}, using the nominal solar cycle simulation results of \citet{pogorelov2009a}, suggested that not only dynamic pressure increases at 1 au, but also 
			the global interactions of the consecutive fast and  slow SW streams alone can produce large-scale compressions of the VLISM.
			This phenomenon can be seen also in the more recent models \citep{fraternale2024b,bera2025}.
			
			However, all earlier models had difficulties in reproducing some details of the recent V1 observations, in particular, a non-decreasing ISMF strength after the shock front, the presence of a magnetic ``hump'', and the persistently strong field strength, which up to date has not recovered its pre-\textit{pf2} values. As discussed by \citet{burlaga2024b}, this gave rise  to the alternative hypotheses about the nature of \textit{pf2} and of the hump region. Those include the possibility that \textit{pf2} is not a wave of solar origin and that it marks the first V1 penetration into the VLISM across the HP \citep{fisk2022}.  On the other hand, \citet{dialynas2024} suggested that V1 had entered a new regime of the VLISM flow, the properties of which are closer to the pristine LISM. Another open question is about the properties of magnetic turbulence, namely, the apparent vanishing of intermittency observed since 2022, which, as demonstrated below,  was incorrectly interpreted as a fact supporting the above-mentioned ideas.
			
			In this study, we use a self-consistent, global, data-driven model of the SW–LISM interaction implemented in the Multi-Scale Fluid–Kinetic Simulation Suite \citep[MS-FLUKSS,][]{pogorelov2008d,pogorelov2014,pogorelov2017a,pogorelov2023iau}, together with a detailed analysis of turbulent magnetic field fluctuations in the Voyager data set currently available to the public,  to demonstrate that the V1 and V2 observations are consistent with a global compression associated with the solar cycle (SC) 24. We also highlight the role of two different LISM magnetic field configurations in the modeled heliospheric dynamic structure and its agreement with the Voyager data.  
			
			We compare the simulation results with the observational data from V1, V2, and New Horizons, and show our model predictions regarding the future observations of shocks/pressure fronts by these spacecraft, including the TS crossing by NH. We discuss the remaining open challenges and limitations in current models.

			The paper is structured as follows. The modeling framework is described in Section \ref{sec:framework}. In Sections~\ref{sec:transientsV1} and~\ref{sec:transientsV2} we discuss the perturbations along the V1 and V2 trajectories, respectively. In Section~\ref{sec:turbulence} we present the turbulence analysis. In Section~\ref{sec:foreshock} we examine the spatial extent of the foreshock associated with the global compression and its potential connection to the ISMF fluctuations recorded by V1. 
			Our conclusions are presented in Section~\ref{sec:conclusions}. Appendices~\ref{sec:appA1} and~\ref{sec:appA2} provide the  details on the global model and  implementation of the boundary conditions, respectively. Appendix~\ref{sec:appB} describes the methodology used for the turbulence data analysis.
			\begin{figure*}[t!]
				\centering
				\includegraphics[width=0.88\textwidth]{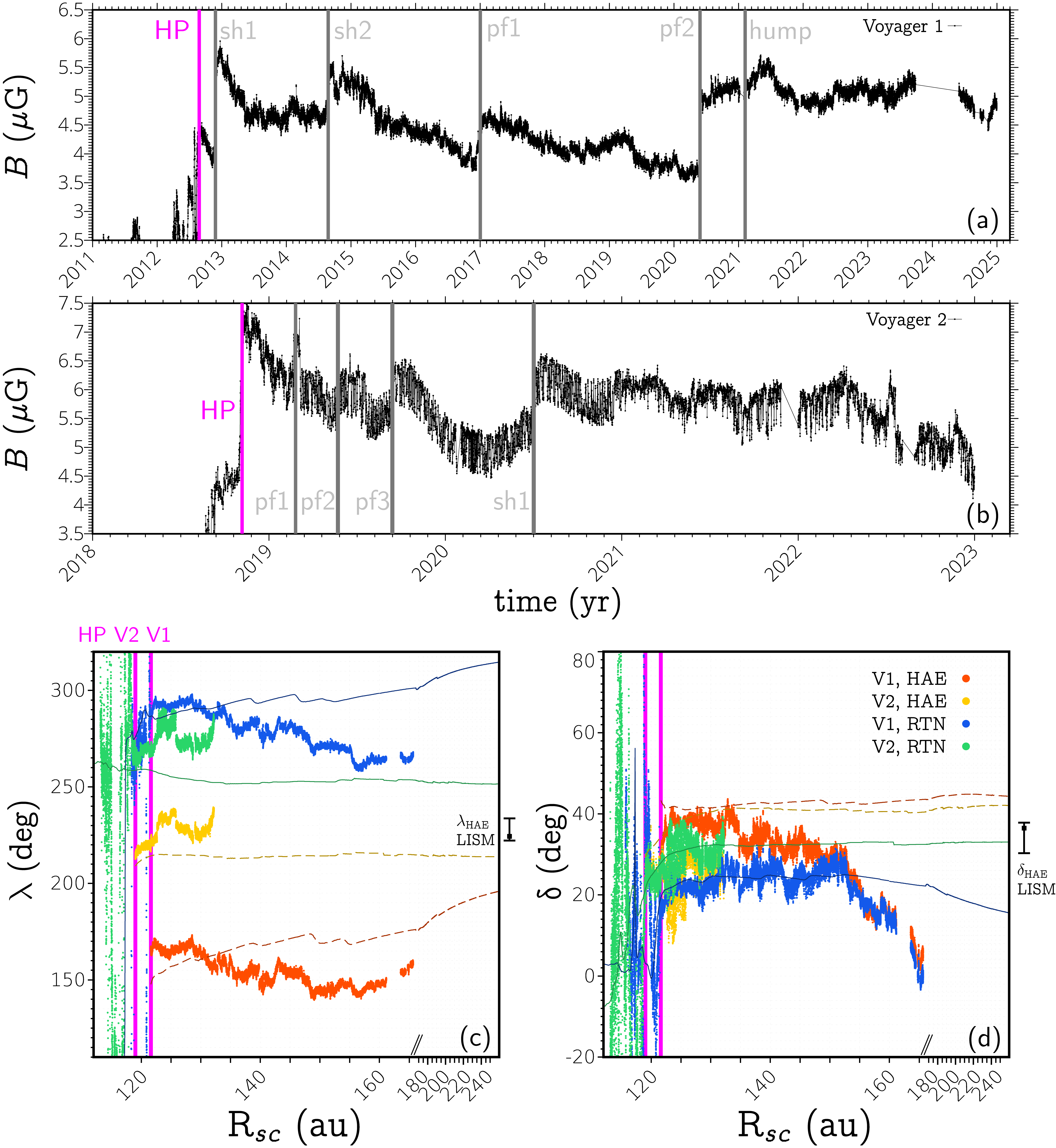} \vskip-5pt  
				\caption{Voyager 1 and 2 MAG data (1 hr averages). Panels (a) and (b) show the magnetic field strength as a function of time at V1 and V2, respectively. Panel (c) shows the azimuthal angles as functions of spacecraft heliocentric distance in both the local RTN coordinate system (green, blue curves) and the ecliptic J2000 (HAE) coordinate system (orange and yellow curves). Using HAE allows for a direct comparison between the two spacecraft.  Panel (d) shows the elevation angles in the same format as panel (c). The solid and dashed curves show the same angles recorded by the spacecraft in simulation \texttt{B} out to 250 au.  On the right side of panels (c,d), we additionally show the range of azimuthal and elevation angles of the far ISMF inferred from the IBEX observations and modeling results by \citet{zirnstein2016}, as well as the values for $B_\mathrm{LISM}=3.5~\mu\mathrm{G}$ used in simulation \texttt{B}, shown by the black square marker. \label{fig:data}}
			\end{figure*}
			
			
			\section{Physical model and simulation framework}\label{sec:framework}
			
			We use the most recent version of our 3D, global model of the SW--LISM interaction implemented in MS-FLUKSS. In particular, for this study we use the multifluid model that includes pickup ions (PUIs) as a separate fluid comoving with the plasma mixture. The flow of the plasma mixture is described by solving the system of ideal MHD equations in the conservation-law form, supplemented by the equations for the PUI density and pressure, as detailed in \citet{pogorelov2016, bera2023a}. The flow of neutral atom  populations is described by the separate systems of Euler gas dynamics equations \citep[see, e.g.,][]{pauls1997,pogorelov2006}.
			In contrast to all other existing models, we use the boundary conditions (BCs) for PUI at the TS, which were derived from the hybrid simulations of \citet{gedalin2023a} (see also the references therein) and first implemented in the global model by \citet{bera2023a,bera2024a}.

			The following modifications are introduced: (i) four populations of neutral hydrogen fluids are used instead of three (we now include the population of H atoms born in the outer heliosheath);  (ii) the mass of helium ions is approximately incorporated; and (iii) the products of charge exchange between protons and neutrals originating in the IHS or the supersonic SW, including PUIs formed in the OHS, are tracked similarly to how it was done in \citet{fraternale2025a}.
			
			Point (ii) is particularly significant, because it leads to significant changes in the solution, and introduces new challenges, some of them being addressed in this paper. While we do not solve separate equations for \HeI, \alp ions or electrons, as is done in MHD/kinetic models, \citep[see, e.g.,][]{fraternale2023,fraternale2024b,fraternale2025a}, the presence of helium ions can be approximately accounted for by adjusting the mass density and pressure in the proton-only model. The feedback of helium ions through charge exchange is not included. However, this effect is of little importance for the present study.

			We have performed four data-driven simulations, with the parameters of the unperturbed LISM summarized in Tab.~\ref{tab:BC}. The boundary conditions are imposed at 1 au and derived from the combination of OMNI, Ulysses, WSO, and Interplanetary Scintillation (IPS) data,  as described in detailed in appendix~\ref{sec:appA2}. We use the most recent model 
			for the latitudinal variation of SW density and velocity in time \citep{porowski2023}. The SW speed is derived from the IPS data, whenever available. Otherwise, the modified methodology proposed by \citet{porowski2024} is used. 
			This data consists of time-latitude maps with the time resolution of a Carrington rotation. This SW model has been used to generate the newest set of IBEX survival probabilities \citep{bzowski2008,mccomas2024} and, to our knowledge, covers the longest time interval available so far for the SW--LISM interaction modeling. These BCs are shown in Appendix \ref{sec:appA2}.
			
			As seen from Tab.~\ref{tab:BC}, cases~\texttt{A} and~\texttt{B} differ only in the magnitude and direction of the LISM magnetic field  \citep[2.93~$\mu$G and 3.5~$\mu$G, respectively, with the corresponding field directions from][]{zirnstein2016}. For each case, we also ran an additional simulation for two solar cycles only (\texttt{A1} and \texttt{B1}, respectively), where the PUI pressure and density equations are used to track also the PUIs born in the OHS.
			
			\begin{deluxetable*}{llccl}
				\tablenum{1}
				\tabletypesize{\footnotesize}
				\tablecaption{Outer boundary conditions and list of performed simulations. \label{tab:BC}}
				\tablewidth{0pt}
				\tablehead{
					Parameter & Units & {Case 1} & {Case 2} & \multicolumn{1}{l}{Description}
				}
				\startdata
				$n_{\p,\lism}$ & \cc & \multicolumn{2}{c}{0.075} & LISM proton density \\
				$n_{\HeI,\lism}$ & \cc & \multicolumn{2}{c}{0.00898} & LISM \HeI density \citep{bzowski2019}\\
				$n_{\HH,\lism}$ & \cc & \multicolumn{2}{c}{0.195} & LISM \HH density \citep{swaczyna2020} \\
				$n_{\He,\lism}$ & \cc & \multicolumn{2}{c}{0.0153} & LISM \He density \citep{gloeckler2004} \\
				$V_\lism$ & km~s$^{-1}$ & \multicolumn{2}{c}{$-25.4$} & LISM bulk flow speed \citep{mccomas2015}\\
				$(\lambda_{V,\lism}, \beta_{V,\lism})$ & deg & \multicolumn{2}{c}{(255$^\circ$.7, 5$^\circ$.1)} & LISM flow direction (Ecliptic J2000) \citep{mccomas2015}\\
				$T_\lism$ & K & \multicolumn{2}{c}{7,500} & LISM temperature (all species) \citep{mccomas2015}\\
				$B_\lism$ & $\mu$G & 2.93 & 3.50 & ISMF strength\\
				$\lambda_{B,\lism}$ & deg & 227$^\circ$.28 & 224$^\circ$.46 & ISMF azimuthal angle (Ecliptic J2000) \citep{zirnstein2016}\\
				$\beta_{B,\lism}$ & deg & 34$^\circ$.62 & 36$^\circ$.61 & ISMF elevation angle (Ecliptic J2000) \citep{zirnstein2016}\\
				\hline
				\multicolumn{5}{c}{\bf Simulations}\\ \hline
				\multicolumn{5}{l}{\texttt{A}\hspace{20pt} Data driven, 5-fluid model, LISM parameters $\rightarrow$ Case 1} \\
				\multicolumn{5}{l}{\texttt{B}\hspace{20pt} Data driven, 5-fluid model, LISM parameters $\rightarrow$ Case 2} \\
				\multicolumn{5}{l}{\texttt{A1}\hspace{20pt} Data driven, 5-fluid + OHS-PUIs model, LISM parameters $\rightarrow$ Case 1} \\
				\multicolumn{5}{l}{\texttt{B1}\hspace{20pt} Data driven, 5-fluid + OHS-PUIs model, LISM parameters $\rightarrow$ Case 2} \\
				\hline
				\enddata
				\tablecomments{%
					Case 1 corresponds to $B_\lism = 2.93~\mu$G and Case 2 to $B_\lism = 3.50~\mu$G. The inner BCs are specified at 1~au and are the same for all runs, described in Fig.~\ref{fig:BCs}.
				}
			\end{deluxetable*}

			\section{Results}\label{sec:results}

			\subsection{Transient disturbances along the V1 trajectory}\label{sec:transientsV1}
			
			Let us focus first on the VLISM disturbances along the V1 trajectory. Since the crossing the HP in August 2012, V1 has recorded two main shock events (hereafter labeled \textit{V1-sh1} and \textit{V1-sh2}) and two pressure fronts (\textit{V1-pf1} and \textit{V1-pf2}) \citep[e.g., see Fig. 1 in ][]{burlaga2023} and our Fig.~\ref{fig:data},  which shows the magnetic field magnitude and direction both in the local RTN coordinate system and in the Sun-centered ecliptic J2000 coordinates. A comparison with one of our models is also provided and will be discussed below. Throughout this study, we adopt the same labeling convention for transients identified along the V2 and NH trajectories. When the context is clear, such as in the figures, we omit the spacecraft prefix. 
			 For all magnetic field analyses presented in this study, we used 48 s Voyager Interstellar Mission MAG data, processed with our own routines, through 2025 June 30 for V1 \citep{dataset-V1-48S-MAG-VIM} and through 2022 December 31 for V2 \citep{dataset-V2-48S-MAG-VIM}.
			
\begin{figure*}[!]
				\centering
				\includegraphics[width=\textwidth]{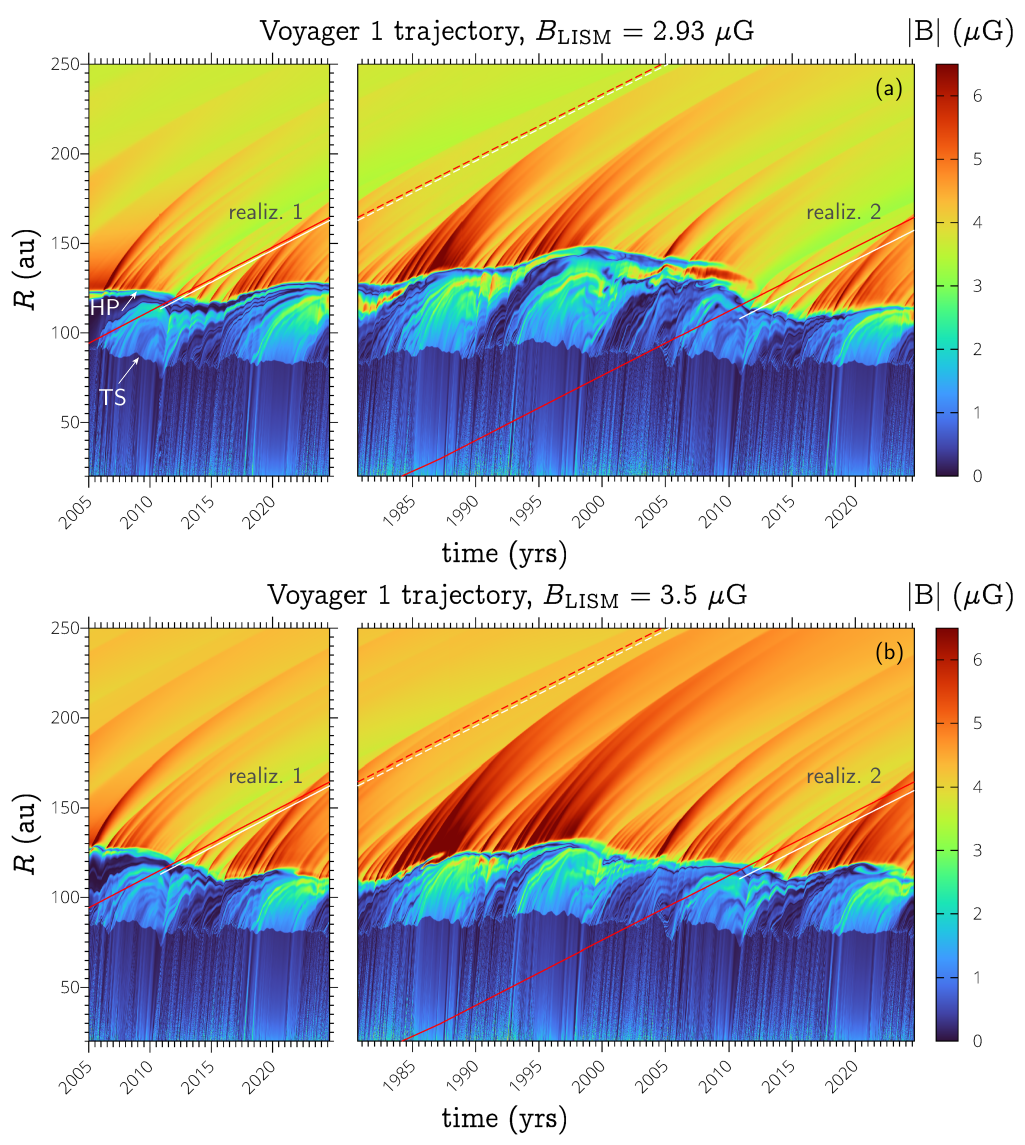} 
				\caption{Space-time distributions of the magnetic field magnitude for simulations A (top panels) and B (bottom panels). The horizontal panel break indicates the moment of time when the BCs at 1 au reach 2024.65 (realization r1) and are restarted from 1980.65 (realization r2). The red line indicates the exact V1 trajectory, while the white curves show the shifted trajectory designed to compensate for the discrepancy in the HP crossing for the two presented realizations. \label{fig:ST_B_MF014_MF015}}
			\end{figure*}

			Figure~\ref{fig:ST_B_MF014_MF015} shows the space-time distributions of the magnetic field strength along the spatial trajectory of V1 over more than four solar cycles, for cases \texttt{A} (panel~a) and \texttt{B} (panel~b), respectively. In both models, data-driven BCs were applied after initializing the simulation with a statistically steady nominal solar cycle solution. The BCs start around 1995 and run until 2024.65. After that, the BCs automatically restart from 1980.65 (details given in App.~\ref{sec:appA2}). 
			Given the high computational cost, this approach allows us to obtain two realizations (hereafter referred to as \texttt{r1} and \texttt{r2}) of the time interval of interest (2012–2025). The second realization accounts for the self-consistently developed, large-scale dynamics of the heliosphere. Overall, we have run six solar cycles. Consequently, the left and right panels of Fig.~\ref{fig:ST_B_MF014_MF015} represent different segments of a single continuous simulation, with the physical time represented on the horizontal axis. 
			
			This kind of plots have been widely used in the literature \citep[e.g.,][]{scherer2003,pogorelov2009a,washimi2011} because they allow one to identify traveling waves, advecting structures, etc., calculate their speeds projected onto chosen  directions, and extract data along the spacecraft trajectory. All presented space-time distributions are constructed using 12-15 day resolution data extracted at runtime due to computational memory constraints. The simulations themselves are performed at a significantly higher time resolution. 
			
			The first observation is that solar activity typically generates one major global perturbation per solar cycle, which creates compressions in the VLISM that can persist for several years and propagate outward beyond 300–400 au from the Sun. The dynamics responsible for the generation of those compressions is more complex than the typical mechanism  of producing shocks driven by the GMIRs such as, e.g., \textit{V1-sh1} or \textit{V1-sh2}. In this case, one should instead pay attention to the global effect that involves (i) the motion of the boundary between the fast and slow SW and (ii) an increase in the dynamic pressure at 1 au, which typically  follows the solar maximum conditions. 
			
			The first mechanism alone may be sufficient to produce weak, 11-year periodic, compressions of the VLISM as suggested by several nominal-cycle simulations in which the dynamic pressure is held constant at the inner boundary \citep[see examples in, e.g., ][]{pogorelov2009a,pogorelov2017b,fraternale2024a,fraternale2024b,bera2025}, or in the solar cycle simulations of \citet[][]{pogorelov2013b} based on Ulysses data.  
			A rapid increase in dynamic pressure that typically occurs each solar cycle after solar maxima \citep[e.g.,][]{sokol2021}, on the other hand, can produce a shock, or pressure front that may later become a shock, in the vicinity of the heliopause, which gains strength due to merging with other shocks propagating ahead of it. However, regular shock merging alone is not sufficient to account for the strong compressions shown in Fig.~\ref{fig:ST_B_MF014_MF015}. This effect was particularly pronounced during solar cycles~21 and~22, leading to large excursions of the HP, as illustrated in Fig.~\ref{fig:ST_B_MF014_MF015}. In model~A, a large-scale instability of the HP is also prominent, with the associated roll structure reaching the V1 latitude in the early 2000s and persisting for a few years.

\begin{figure*}[t!]
				\centering
				\includegraphics[width=\textwidth]{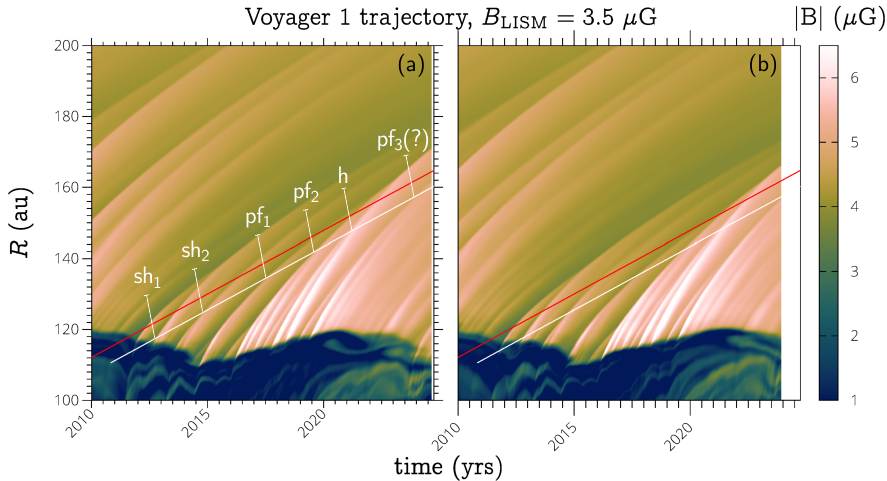}
				\caption{Details of the magnetic field magnitude distribution, during the time interval from 2010 to 2025, as obtained from simulations \texttt{B} and \texttt{B1} with identical BCs. \label{fig:ST_B_MF015pui}}
			\end{figure*}

\begin{figure*}[h!]
				\centering
				\includegraphics[width=0.85\textwidth]{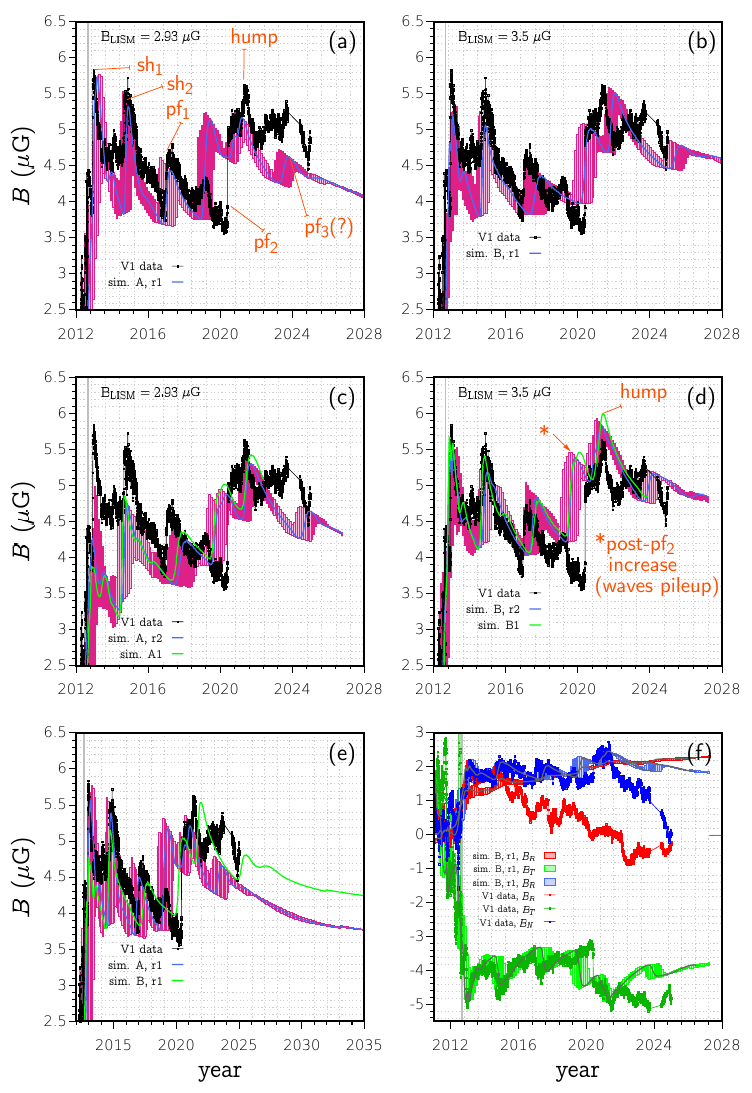}\vskip-10pt
				\caption{Interstellar magnetic field distributions at V1: comparison between the simulation results and V1 data. Panels (a) to (d) show the magnitude distributions for the different cases and different realizations of the 2012-2025 interval within each case. Panel (e) compares results of simulation \texttt{A} and simulation \texttt{B}, highlighting the long-term evolution. Panel (f) shows the individual components of the ISMF and their comparison with observational data. In each panel, the error bars represent the uncertainty associated with a $\pm 1.5$ variation in the V1 trajectory used for data extraction, reflecting the uncertainty in the HP position, while the solid curves show the simulation data extracted along the shifted trajectory shown in Fig. \ref{fig:ST_B_MF014_MF015} (white line).\label{fig:linear_B}}
			\end{figure*}

			The major result of this study is that this dynamical mechanism can explain the nature of the \textit{V1-pf2} event (2020.4), and the relatively strong magnetic field observed to date. Figure~\ref{fig:ST_B_MF014_MF015} illustrates the encounter of a strong compression wave with V1 in models. The red line shows the actual spacecraft trajectory, while the white line corresponds to a shifted trajectory that compensates for the discrepancy in the heliopause crossing, because our simulations systematically underestimate the heliosheath thickness by a few au. Interestingly, overestimation was more common in many previous studies. Therefore, for a meaningful comparison with the Voyager data in the LISM, it is necessary to identify the HP crossing and to correctly associate the individual observed and simulated shocks, and in this way determine an appropriate spatial shift. At V1, such a shift is  $\sim$2.2 au in realization \texttt{A-r1}, $\sim$7 au in realizations \texttt{A-r2} and \texttt{A1}, $\sim$3.5 au in realization \texttt{B-r1}, and $\sim$4.8 au in realization \texttt{B-r2}  and simulation \texttt{B1}).  This is better seen in the zoomed view of Fig.~\ref{fig:ST_B_MF015pui}, which shows the same time interval in simulations  \texttt{B} and  \texttt{B1}. If the simulation did not underestimate the heliopause distance by a few au, the first shock observed would correspond to the one labeled as \textit{sh1} in Fig.~\ref{fig:ST_B_MF015pui}. This identification was performed manually for all cases presented in this study. It goes without saying that the line shift was only used for VLISM analysis. 
			We also highlight the role of an intrinsic stochasticity in the solution, and the need for uncertainty quantification, similarly to what is regularly done in inner heliospheric simulations \citep[e.g.,][]{hegde2025}. Identical BCs may lead to different VLISM distributions as a result of variations in the heliopause structure, which is corrugated by Kelvin-Helmholtz and Rayleigh-Taylor–type instabilities 
			\citep[e.g., ]{fahr1986,ruderman1993,ruderman1995,zank1996b,zank1996c,florinski2005,borovikov2008,florinski2015}, as well as by magnetic reconnection as shown, e.g., by 
			\citet{pogorelov2017b},  where  the first systematic numerical investigation of these instabilities was also performed in 3D. The corrugated heliopause surface and its effect on the geometry of the interstellar shocks in the meridional plane are shown in the top panels of Fig.~\ref{fig:foreshock_streamline}. 
			The time distributions of $|\mathbf{B}|$ observed by a virtual V1 spacecraft traveling through the simulation domain are shown in Fig.~\ref{fig:linear_B}, alongside with the observational data (black points).

			
			The error bars represent the excursion of the solution resulting from the shifts of the virtual spacecraft trajectories by 
			$\pm$1.5 au. This provides a rough proxy for the sensitivity of the solution (particularly, the shock positions) to the uncertainty in the HP location.

			Our results can be summarized as follows.
			
\begin{figure*}[t!]
				\centering
				\includegraphics[width=0.9\textwidth]{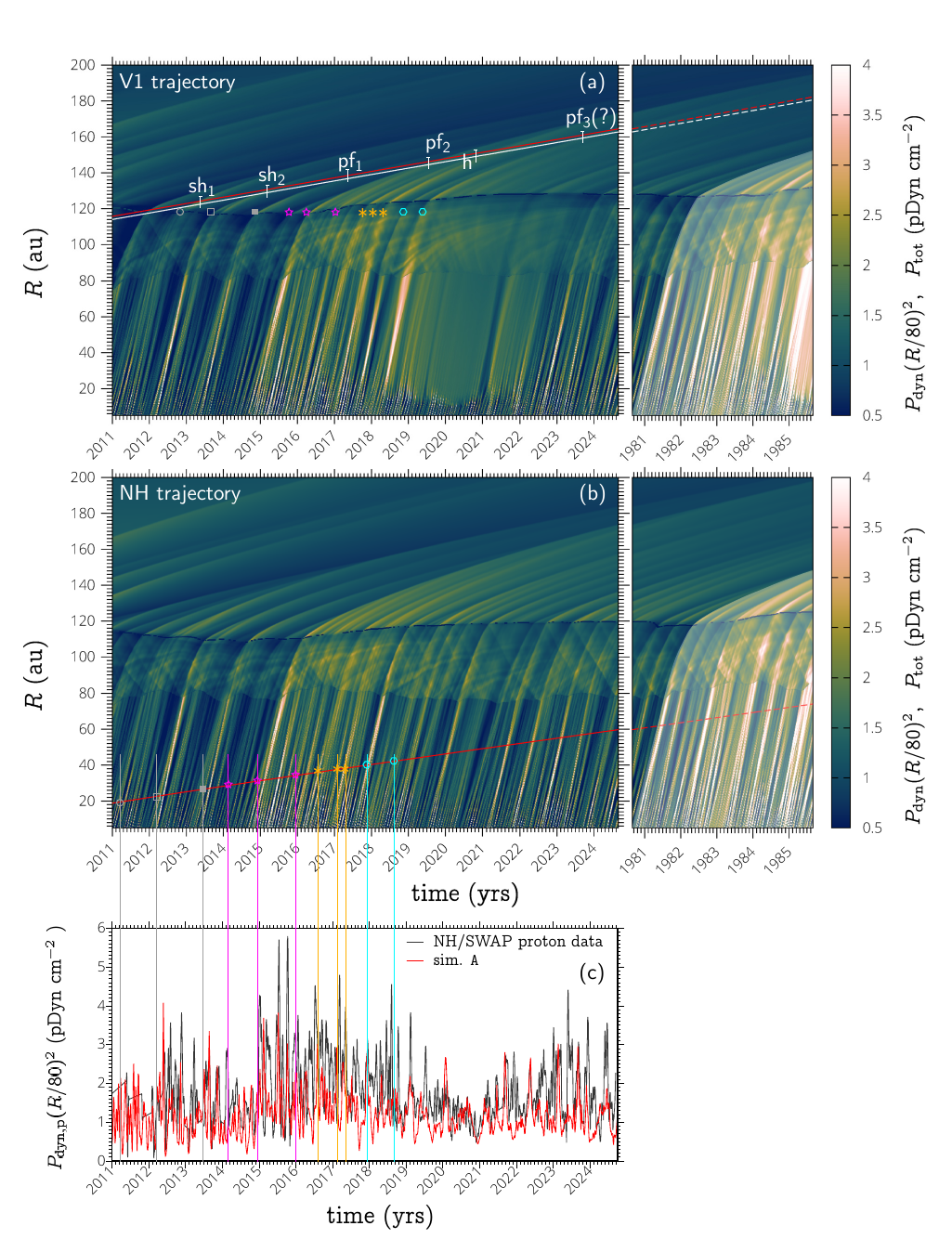} 
				\caption{Tracing the origin of interstellar compressions back to the supersonic SW. Panels (a) and (b) show the space-time distributions of total pressure (for $R>R_{TS}$ only) and the scaled dynamic pressure, $p_\mathrm{dyn} (R/80)^2$ (in the supersonic SW region ($R<R_{TS}$) along the V1 and the NH trajectories, respectively. Panel (c) shows the scaled dynamic pressure of protons measured by NH (black points, Carrington-rotation averaged data) together with the simulated one. The perturbations that result in the interstellar compressions seen by V1, including the ``hump'' (h), are identified along the NH trajectory and labeled with different symbols.
					\label{fig:shock_tracing_V1} }
			\end{figure*}

			\begin{itemize}
				\item \textbf{The nature of \textit{V1-pf2}.} As shown by \citet{pogorelov2021} and \citet{zirnstein2024}, both of which
				analyzed the simulations of \citet{kim2017b}, 
				\textit{V1-pf2} may be due to merging of the three shocks, with the dominant contribution coming from the shock driven by the increase in dynamic pressure at 1~au that occurred around 2015. In our simulations, however, in contrast to \citet{kim2017b}, \textit{pf2} corresponds to the merger of two shocks when this pressure front reached V1. Another (the third) compression is delayed in all our models and creates the ``hump'' (see below). Regardless of these differences, we emphasize that \textit{pf2} is not a regular GMIR-driven shock, but is rather the product of a stream interaction cased by the solar-cycle.
				\item \textbf{Magnetic field behind \textit{V1-pf2}}. Most of our simulations underpredict the arrival time of \textit{pf2} when matching the encounters with \textit{sh1} and \textit{sh2}, consistent with the findings of \citet{kim2017b}, with the exception of simulation~\texttt{B-r1}. For the first time, we were able to reproduce a shock/compression with a non-decreasing $B$ immediately behind it, as observed by the moving spacecraft (see realizations~\texttt{A-r2}, \texttt{A1},  \texttt{B1} in panels~(c,d) of Fig.~\ref{fig:linear_B}). This feature has been particularly challenging to capture, and distinguishes \textit{V1-pf2} from the earlier shocks. Typical VLISM shocks exhibit a jump-ramp profile, or an N-wave structure \citep{whitham1974}, associated with the rarefaction that occurs when the heliopause retreats and the shock propagates outward in an approximately spherical geometry. Such magnetic field profiles are ubiquitous in the VLISM, even for  minor perturbations, as discussed by \citet{fraternale2021a}. First, the global compression following \textit{V1-pf2} provides conditions for a shallower ramp. Additionally, we demonstrate that the sustained increase in $B$ observed immediately behind \textit{pf2}, and lasting up to the arrival of the ``hump'', is due to the presence of multiple waves generated at the heliopause between 2016 and 2018 and their merging, or ``pileup'',  as clearly seen from Fig.~\ref{fig:shock_tracing_V1}. While a perfect agreement between the simulations and data is not expected, because of the uncertainty in BCs and resolution limitations, this is the first time that a non-decreasing $B$ profile behind the shock has been reproduced, as indicated in panel (d) of Fig.~\ref{fig:linear_B}.
				\item \textbf{The ``hump'' feature}. In our models, the ``hump'' in the data (indicated in Fig.~\ref{fig:linear_B}a) corresponds to an additional compression wave that exists in the models (see Fig.~\ref{fig:linear_B}d and the perturbation labeled as~\textit{h} in Fig.~\ref{fig:shock_tracing_V1}). It exhibits a bell-shaped profile, which is approximately reproduced in simulations \texttt{B-r2} and \texttt{B1} as a consequence of ongoing merging with the higher-frequency waves propagating between \textit{V1-pf2} and the hump itself. More details are provided below.
				
			\end{itemize}

			We have attempted to trace these perturbations back to the supersonic SW. For this purpose, it is useful to examine the total pressure ($p_\mathrm{tot}=p_\magn+p_\tth$) in the inner and outer heliosheaths, as most of the total pressure is transmitted across the HP when a nearly-perpendicular shock hits it.  In the supersonic SW, by contrast, the dynamic pressure is dominant. Therefore, for the purpose of tracing disturbances, in Fig.~\ref{fig:shock_tracing_V1} we combine these two quantities in a single plot. We show the scaled dynamic pressure, $p_\dyn (R/80 au)^2$, in the supersonic SW and the total pressure elsewhere, along the V1 and NH directions (panels a and b). 
			In the IHS, forward-propagating shocks originating at the TS are clearly visible in all space–time distributions, along with advecting fluctuations associated with entropy perturbation  modes \citep[e.g.,][]{borovikov2012}. Upon reaching the HP, part of these shocks is reflected and propagates sunward, where it interacts with and affects the TS motion. This is also seen in the PUI quantities shown in Fig.~\ref{fig:ST_PUI_V1}.
			
			Different interstellar shocks and compressions in the data and in the model are labeled for clarity in Fig.~\ref{fig:shock_tracing_V1}. Question marks indicate the shocks predicted by the model, but not yet observed. In panel (c), the dynamic pressure of protons measured by the NH/SWAP instrument, at the 1-day resolution \citep{dataset-NH-SWAP-SW}, is shown together with the simulation results. 
			
			
			Colored symbols identify the major perturbations that, through their  evolution and merging, lead to the formation of the shocks and pressure fronts observed by V1 in the VLISM. In particular, it is shown that the ``hump'' itself is formed by merging of three main compressions possibly observed by NH between mid-2016 and mid-2017 (yellow lines and symbols). Moreover, in all simulations a pressure front reaches V1 in 2024 (tagged as \textit{V1-pf3(?)}), but this event may have coincided with a large gap in observational data. Interestingly, signatures of enhanced fine-scale turbulence were observed ahead of this data gap (see Sec.~\ref{sec:turbulence} and Fig.~\ref{fig:spectraV1finescale}), indicating that a shock may indeed have come across the spacecraft during that time.
			
			It should be noted that  the magnetic field strength in all models does not recover to its pre-\textit{pf2} values ($\sim 4 \mu$G) until at least 2030. So far, this behavior is consistent with the observations. We suggest that this is due to the global compression associated with SC-24. Models that adopt $B_\lism = 3.5,\mu$G appear to agree better with the magnetic field strength observed by V1, as the background negative radial gradient of $B$ is weaker in this case (Fig.~\ref{fig:linear_B}e). Future data by V1 may in fact help constrain the pristine LISM conditions, with the aid of global models.
			
			Panel (f) of Fig.~\ref{fig:linear_B} highlights the behavior of the components of \textbf{B}. The disagreement between the trends in the simulated and observed radial component, especially after $\sim$2015, has been the subject of recent discussions, see  \citet{rankin2023} and the Appendix of \citet{burlaga2024a} (c.f. also the magnetic field directions shown in Fig.~\ref{fig:data}c,d). Despite our efforts to reproduce both large-scale compressible and transverse fluctuations in our models, this discrepancy remains. Current models are consistent with all of our simulations (steady-state as well as nominal solar-cycle runs) in terms of the large-scale trend exhibited by $B_R$. We also note that the fluctuations in $B_R$ are in general underestimated by the model, compared with the fluctuations in $B_N$. Interestingly, V1 recorded a rapid decrease in $B_N$ after 2021, which is stronger than the one produced by the models.  The corresponding rotation of $\mathbf{B}$ is seen even more clearly in the magnetic field vector angles shown in Fig.~\ref{fig:data}(c,d). Clearly, the angles in heliocentric ecliptic coordinates at V1 and V2 tend toward the common asymptotic value, although the convergence is slow, different for the two directions, and is not reached within $\sim$500 au. Only in the recent V1 data, the azimuthal angle appears to be rotating towards the expected value, while the elevation angle is evolving in the expected direction but too rapidly, as compared to all models. Although Voyager 2 is likely to be insufficiently far from the heliopause to make a meaningful extrapolation, the  modeling results at V2 seem to underestimate the azimuthal angle while reproducing the elevation angle quite well.
			
			Finally, our models show that the speed of propagation of solar-cycle-related perturbations decreases with distance, eventually reaching speeds comparable to that of V1 (Fig.~\ref{fig:ST_B_MF014_MF015}). Therefore, even if V1 were to remain operational out to distances of 200–250 au, it would hardly be overtaken by the global transient perturbation associated with the current solar cycle, SC~25. We predict that V1 will record relatively quiet VLISM conditions in the coming years, in terms of solar-driven perturbations. The situation is quite opposite for V2, as will be discussed below.
			
			\subsection{Transient disturbances along the Voyager 2 trajectory}\label{sec:transientsV2}
			
			Voyager 2 crossed the heliopause in November 2018 at a distance of 119.0 au and, up to 2022, observed at least two major compression waves and a shock, as detailed in \citet{burlaga2022}. Interestingly, the magnetic field profiles of shocks and pressure fronts observed by V2 differ from those observed by V1, with the former appearing as smoother, wave-like structures.  Moreover, when V2 crossed the HP, it recorded remarkably high magnetic field strengths, which have been attributed to its location in the southern hemisphere, where the outer heliosheath plasma is more compressed. Finally, V2 did not observe a strong shock corresponding to \textit{V1-pf2}. To date, no modeling effort has been able to fully explain these observations in their entirety. Here we propose possible explanations based on the current models.

			\begin{figure*}[h!]
				\centering
				\includegraphics[width=0.95\textwidth]{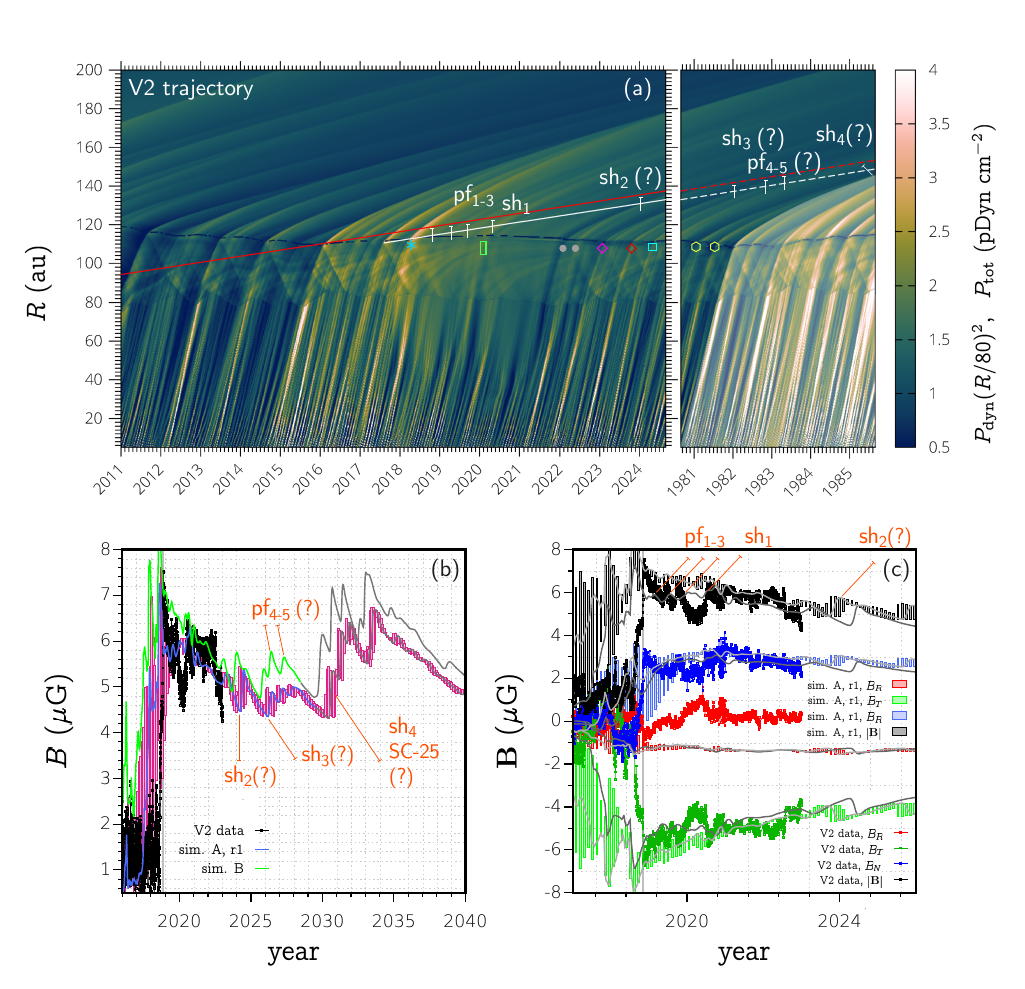} \vspace{-10pt}  
				\caption{Space–time evolution of the total and dynamic pressures along the V2 trajectory, defined as in Fig.~\ref{fig:shock_tracing_V1} (panel a), and comparison of the simulated V2 temporal distributions of the magnetic field strength (panel b) and vector components (panel c) with V2 observations.
					\label{fig:shock_tracing_V2} }
			\end{figure*}
			Our modeling results extracted as a function of time along the V2 trajectory for the 2011–2029 time interval are summarized in Fig.~\ref{fig:shock_tracing_V2}. Panel (a) shows the combined total and dynamic pressure space–time distribution from simulation \texttt{A}, constructed in the same way as in Fig.~\ref{fig:shock_tracing_V1}, while panels (c) and (d) show the extracted linear distributions of the magnetic field along the moving spacecraft trajectory, from both simulation \texttt{A} and \texttt{B}. Our main findings can be summarized as follows. 
			
			\begin{itemize}
				\item Reproducing V2 data is more challenging than reproducing V1 data; this is a well known issue in the heliospheric modeling community. In particular, all models presented here underestimate the heliopause standoff distance at the time of the crossing by $\sim $3.5~au, $\sim$4.5~au (realization~\texttt{A-r1}), $\sim$7.2~au (realization~\texttt{A-r2} and \texttt{A1}), and up to $\sim$8.5~au (for \texttt{B} and \texttt{B1}). As discussed in Section~\ref{sec:transientsV1}, we believe that this issue can be addressed in future simulations using MHD-plasma/kinetic-neutrals models, and by improving the inner BCs. Here, as was done for V1, a meaningful comparison of transient structures in the VLISM requires  proper shock identifications and application of a spatial shift to the virtual spacecraft trajectory (see the white line in Fig.\ref{fig:shock_tracing_V2}a). This once again demonstrates that a global perspective is necessary for the interpretation of spacecraft observations. In this regard, we also note that the interstellar plasma velocity in the models remains radially outward throughout the entire interval up to 2021, as shown in Fig.~\ref{fig:ST_Ucomps_V2}, which is due to the global effect associated with SC-24. 
				\item The model reproduces the presence of weak compression waves with smooth profiles (\textit{pf1-pf3} in Fig.~\ref{fig:shock_tracing_V2}), as well as a stronger wave that we associate with \textit{V2-sh1}. However, the amplitude of these compressive waves is lower in the models than it was observed. Tracing the perturbations back to the SW regions, the model suggests that the dynamics may be significantly influenced by the location of the fast/slow SW boundary during the 2017.5--2021 interval (the declining phase of SC-24; see Fig.~\ref{fig:BCs}). 
				It is worthwhile to note that high-cadence OMNI data is used as BCs only in the sector region. The properties of perturbations in the fast SW are therefore not well constrained, which may be an explanation for the perturbations modeled VLISM to appear smoother than those observed.
				\item One of the questions raised by the heliospheric community is about why V2 did not observe the perturbation corresponding to the observed \textit{V1-pf1}. A straightforward answer is that the SC-24 perturbation of the VLISM was generated in the VLISM before V2 crossed the HP. This is clearly visible in all plots.
				\item Magnetic field strength along the V2 trajectory shows that three factors contribute to the observed values of $B$ around $7\,\mu$G. As previously noted, magnetic pressure is higher at the V2 latitudes than at the V1 latitudes, provided that one uses the IBEX-compatible $\mathbf{B}_\lism$ directions in steady-state models. However, this is not sufficient for our models to reproduce field strengths of $\sim 7,\mu$G. 
				In fact, data-driven models indicate that the global perturbation associated with \textit{V1-pf2} further contributes to enhancing the magnetic field observed along the V2 trajectory, as V2 travels behind this perturbation (and V2 will never be able to reach the shock front, as it travels more slowly). Ultimately, the best fit to the V2 data is obtained when placing the virtual V2 crossing of the HP immediately after a shock (identified by the cyan star symbol in Fig.~\ref{fig:shock_tracing_V2}a). This configuration also explains the declining magnetic field strength observed after 2019. Interestingly, this shock can be traced back to the same perturbation that in the models generates the magnetic hump observed along V1.
				\item Magnetic field components, shown in Fig.~\ref{fig:shock_tracing_V2}(c), are in reasonable agreement with the data. As for V1, the major discrepancy is in its radial component. Near-perpendicular fluctuations ($\delta B_R, \delta B_N$) are the most challenging to reproduce.
			\end{itemize}
			
			As far as the model predictions are concerned, V2 may observe a relatively strong shock or pressure front in 2024$\pm0.5$~yr, labeled as \textit{sh2(?)} in Fig.~\ref{fig:shock_tracing_V2}. It is currently unclear whether this perturbation has already been recorded by the magnetometers, as published data extend only up to 2023.0. This shock may result from merging of multiple shocks and structures observed by NH during the 2021.5–2022.5 interval \citep[see Fig.~1 in][]{shrestha2025}.
			
			A second shock (\textit{sh3(?)}) may cross V2 by 2026; we trace this event back to the increase in dynamic pressure and the shocks recorded by NH between 2022.5 and early 2023. \textit{Sh3} is followed by two additional compressions, the last of which is expected around 2028. After 2029, the modeling results along the V2 trajectory depend on BCs that restart from 1980.65. As a result, quantitative predictions beyond this time are not meaningful, which is why the corresponding curves are shown in gray and the region on the right-hand side of panel~(a) is shaded in white.
			
			Of major interest are the long term predictions. While V1 is expected to record relatively quiet conditions, this is not the case for V2. Our simulations indicate that each solar cycle produces a global response in the form of one or multiple shocks and a large-scale compression of the VLISM. If a significant increase in dynamic pressure occurs after solar maximum in SC-25 as well, it is likely that V2 will observe a major event, of the same nature of \textit{V1-pf2} and similar to \textit{V2-sh4(?)} shown in Fig.~\ref{fig:shock_tracing_V2}(b), expected around 2030–2031.

			While this study is focused on the outer heliosheath modeling, as already shown in Fig.~\ref{fig:shock_tracing_V1} the perturbations originate in the SW. Therefore, we would like to show how the models result compare with V2/PLS and NH/SWAP data in the supersonic SW, and make a prediction for the crossing of the TS by NH.
			
\begin{figure*}[t]
				\centering
				\includegraphics[width=\textwidth]{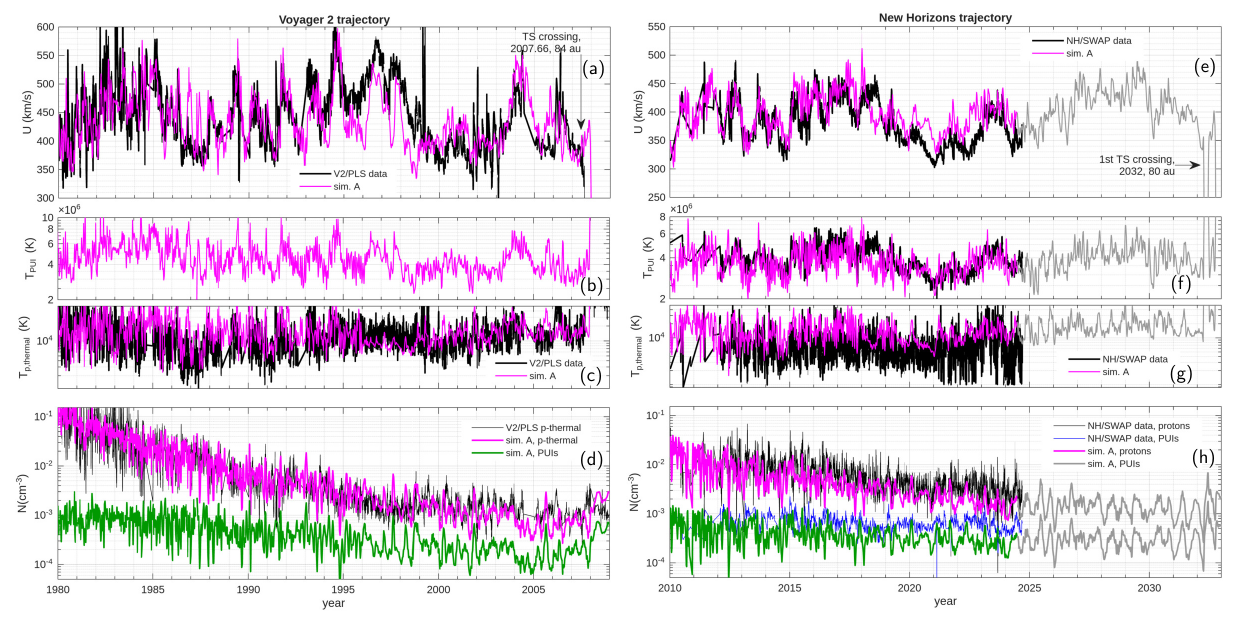} 
				\caption{Linear distributions in the supersonic SW along the V1 (left panels) and NH (right panels) trajectories from simulation \texttt{A}  extracted along the moving-spacecraft trajectory, and a prediction for the TS crossing by NH. From the top to the bottom, panels show the SW speed, PUI and core SW temperature, and the density distributions.  The simulation results are compared with observations from Voyager/PLS and NH/SWAP at the 1 day resolution. }\label{fig:TS_NH}
			\end{figure*}

			Figure~\ref{fig:TS_NH} shows, in the left panels from top to bottom, the speed, temperature, and density distributions along the V2 trajectory. Here use the V2/PLS data are available  at 192 s resolution, averaged to the 1 day resolution  \citep{dataset-V2-PLS-HiRes}. 
			
			In the right panels, The same quantities are shown along the NH trajectory.  Here we used the NH/SWAP SW dataset  \citep{dataset-NH-SWAP-SW}, and PUI dataset \citep{dataset-NH-SWAP-PUI}. The middle panels display the temperatures of the core (thermal) and PUIs. No time or spatial shift is applied to the virtual spacecraft trajectories. 
			 Note that the simulated results were extracted along the moving spacecraft trajectories at runtime with an output cadence of about 12 days, although the underlying simulation has temporal resolution higher than 1 day and grid size of 0.1 au. Therefore, there is an apparent smoothing of the simulation results in Fig. \ref{fig:TS_NH}, which is a visualization effect.

			The modeled SW speed generally shows the best agreement with the data \citep[as in][]{kim2016}, although discrepancies between the models and the NH/SWAP observations are present, particularly during the time intervals 2018.5–2019, 2020–2021, and 2022-2023. This suggests that further improvements to the inner boundary conditions are required. Remarkably, the PUI 
			temperature is very well reproduced. The ``bump'' in PUI temperature observed by NH between 2015 and 2019 is a dynamical effect associated with the solar cycle. Moreover, when smaller-scale pressure pulses or shocks are present in time-dependent models, there is no need for an ad hoc heating term for PUIs in the equations.
			
			A puzzle still remains regarding the radial proton density trend along the NH trajectory. While the models reproduce the density along V2 reasonably well, the density is underestimated by almost a factor of two at 60 au, along the NH trajectory. This is surprising, given that the inner BCs are based on OMNI data in the ecliptic, therefore higher uncertainties would instead be expected at higher latitudes. The PUI density at NH is underestimated even more, as this bias reflects the combined effects of an underestimated proton density on the one hand and an overestimated neutral H filtration at the HP on the other. The latter effect is not a major concern because it can be resolved by using MHD-plasma/kinetic-neutrals models, as suggested by our recent kinetic simulations, as well as by earlier comparisons between kinetic and multifluid simulations,  as shown in Fig. 4 of \citet{alexashov2005} and  Fig. 5 of \citet{pogorelov2009c}.
			It is also noteworthy that the heating rate of the core SW protons by PUIs is assumed to be proportional to the charge exchange source term source in the pressure equation for PUIs. However, here we adopt a  3.75\% scaling factor, which is less than  5\% used in \citet{bera2025}. 
			This provides a good representation of the core proton temperature along V2, but leads to an overestimation at NH.  A better agreement with the core proton temperature at NH was obtained in the steady-state MHD/kinetic models by \citet{fraternale2023}, who used a different heating model based on the theory of PUI-generated waves.
			Thus, to address this discrepancy one would require involvement of a proper turbulence transport model, which goes beyond the scope of this study. Interestingly, we needed a smaller fraction of the PUI energy to be transferred to the core protons for the simulations with high-cadence BCs, as compared to the steady-state models. 
			This is consistent with the results of \citet{korolkov2022}, where it was noticed that highly variable BCs can provide substantial heating to the core protons even if the turbulence effects are disregarded. Those authors attributed this phenomenon to heating by multiple shocks in the SW. In principle, one can attribute this also to the numerical heating due to the ``turbulent'' character of the flow. Nevertheless, it is clear that numerical effects can only be considered as partially fulfilling the role of actual SW turbulence, which require one to solve the Reynolds-averaged MHD equations accompanied with turbulence models, as, e.g., in \citet{kryukov2012,usmanov14}.
			
			The TS crossings by the two Voyager spacecraft are reproduced with good quantitative accuracy, at the level of less than 3\% in heliocentric distance.  For V1, the observed crossing at 2004.959 (94.0 au) occurs later and at a larger heliocentric distance than predicted by all simulations, which yield crossings at 2004.29 (91.6 au; \texttt{B-r2}) and 2004.51 (92.38 au; \texttt{A-r2}). This corresponds to an inward bias of about 1.6–2.4 au.

			In contrast, for V2 the observed crossing at 2007.668 (83.7 au) lies within the simulated range spanning 2006.86–2007.93, with corresponding TS distances of 81.1 au (\texttt{B-r2}) and 84.48 au (\texttt{A-r2}), respectively. In addition, simulation A exhibits a realization-1 crossing at 2007.41 (82.85 au; \texttt{A-r1}), differing by about 1.6 au from the realization-2 result. Differences are also found between models employing the same boundary conditions: the V2 crossing from \texttt{B-r2} and \texttt{B1} differ by 0.6 years and by 0.1 year from \texttt{A} to \texttt{A1}. This spread reflects  the sensitivity of the TS location to intrinsic variability of the IHS dynamics. Overall, simulation \texttt{A} provided the best agreement with the Voyager observations.
			
			For NH, the gray parts of the curves in Fig.~\ref{fig:TS_NH} show the model results obtained when the BCs after 2024.65 are continued using the declining phase of SC-21 as described in Appendix \ref{sec:appA1}. Under this assumption, NH crosses the TS for the first time in 2032.27 in simulation \texttt{A-r1}, at a heliocentric distance of 81.46 au. Three crossings are predicted, with the last occurring in 2032.77 at 82.88 au. In simulation \texttt{B}, the first crossing occurs earlier, at 2031.0 (77.8 au), with the last crossing at 2031.62 (79.6 au). These predictions should be regarded as an upper limit, since SC-21 was stronger than the current solar cycle. Notably, the predicted TS distance for NH is close to that obtained from the nominal solar-cycle MHD/Kinetic models of \citet{fraternale2024b}, who find a TS location near 80 au (see their Fig.~1b).


			\subsection{Turbulence and intermittency}\label{sec:turbulence}
			Most of the compressible fluctuations observed in situ by Voyager in the OHS \citep{burlaga2015,burlaga2018} are believed to result from SW fluctuations transmitted into the VLISM though the HP motion \citep{zank2017}. Compressible, fast-magnetosonic modes are expected to be damped relatively quickly with distance or be converted into transverse fluctuations \citep{zank2019}. However, the situation is complicated by the presence of shocks, temporal variability, the complex, non smooth structure of the HP, and small-scale instabilities that may generate local compressible fluctuations \citep[see the review by][]{fraternale2022ssr}. The interpretation of spacecraft observations cannot rely on the Taylor hypothesis and requires considering the motion of the spacecraft through a nonuniform background plasma with varying bulk velocity (see Figs.~\ref{fig:ST_Ucomps_V1} and \ref{fig:ST_Ucomps_V2}).
			
			The presence of intermittency is a characteristic of turbulence.  Intermittency in magnetic field fluctuations in the VLISM was identified in situ Voyager 1 data by \citet{fraternale2019a}, who analyzed several time intervals up to 2016.67 and spacecraft-frame frequencies below $2.8\times10^{-4}$~Hz using structure-function analysis. Subsequently, \citet{fraternale2020a} discovered fine-scale intermittent fluctuations at frequencies up to $6.9\times10^{-3}$~Hz ahead of the shock \textit{V1-sh2}, possibly associated with kinetic processes occurring in its foreshock region. This was a rather unique event, as no other interstellar interval exhibited non-Gaussian statistics at such small scales. \citet{burlaga2020} highlighted the presence of weak intermittency in several V1 and V2 time intervals using hourly averaged data for the analysis.  
			\citet{fraternale2021a} further investigated the nature of intermittency, highlighting the role of steepened waves in producing non-Gaussian statistics, as well as the strong variability of these statistics with time scale and with the chosen observational interval. In that study, we also pointed out the presence of fine-scale intermittency in data starting from late 2017–2018, but without investigating its origin, which we attempt to do here. More recently, \citet{burlaga2023} investigated intermittency at V1 during 2022 and 2023, and \citet{burlaga2024b} focused on 2023. 
			\citet{khoo2025} investigated intermittency using 1~hr magnetic-field increments and a method based of fitting using kappa distributions, confirming that apparent Gaussian behavior can arise from limited statistics and that residual intermittency persists in the 2023 data.

			One apparently puzzling aspect of the most recent V1 observations is the reported  notable drop in intermittency following the magnetic ``hump'' region \citep[observed in 2022; see][]{burlaga2024a,burlaga2024b}. This behavior was incorrectly described as being unlike any previous VLISM observations and led some authors to suggest that V1 may have entered a new region of the LISM, or that it may not have been in the LISM prior to its encounter with \textit{V1-pf2}. Here we show that neither the turbulence observations nor global models support these conclusions.
			
			Due to the peculiar distribution of data gaps of a few hours per day, we have performed a separate analysis of large-scale and fine-scale fluctuations.
			

			\begin{figure*}[t]
				\centering
				\includegraphics[width=0.9\textwidth]{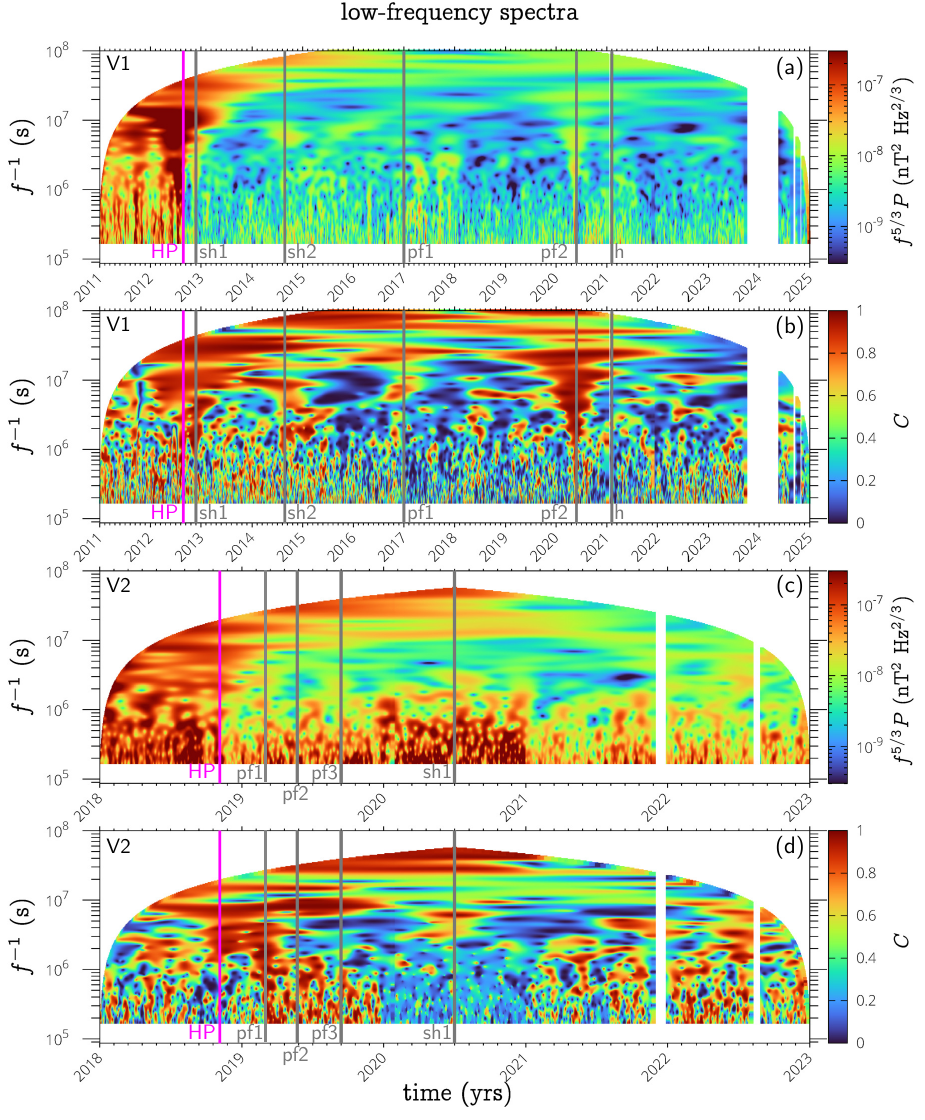}  
				\caption{Wavelet-based power spectral density (PSD) of magnetic field fluctuations measured by V1 and V2 in the VLISM. Panels (a) and (b) show the compensated total PSD $P=P[B_\parallel]+P[B_{\perp1} + P[B_{\perp2}$ (trace) and the magnetic compressibility, $C=P[B_\parallel]/P$, at V1. The same quantities are shown in panels (c) and (d), respectively, for V2. The HP crossing is indicated with a  vertical magenta line, while the gray lines indicate the major shocks or pressure fronts\label{fig:spectraLargescale}.}
			\end{figure*}

			We conducted an analysis of large-scale fluctuations at both V1 and V2, corresponding to turbulence observed over frequencies from $10^{-5}$~Hz to $10^{-8}$~Hz, using MAG data resampled at a 1-day cadence. The analysis is based on a wavelet technique following the method described by \citet{bowen2020}, which performs a field-aligned decomposition of the fluctuations. We used a Morlet wavelet transform, with the mother wavelet defined as
			\begin{equation}
				\psi(t,s) =
				\frac{1}{\sqrt{\sigma_s}\,\pi^{1/4}}
				\exp\!\left(-\frac{t^2}{2\sigma_s^2}\right)
				\exp\!\left(- i\,\frac{2\pi t}{s}\right),
			\end{equation}
			where $s$ is the scale and the Gaussian width is given by
			$\sigma_s = {n_{\mathrm{o}}\, s}/{2\pi}$ and $\omega_0 = 2\pi$. The number of effective oscillations under the Gaussian envelope is set to $n_{\mathrm{o}}=8$, providing enhanced frequency resolution at the expense of temporal resolution, which we preferred for inertial-range turbulence analysis.
			
			Each numerical wavelet is normalized to unit energy,
			\begin{equation}
				\int |\psi(t,s)|^2\,dt = 1,
			\end{equation}
			and the scale-to-frequency conversion is given by
			$f = {\omega_0}/({2\pi s\,\Delta t})$,
			where $\Delta t=24$ hr is the sampling interval.
			
			For the projection into field-aligned coordinates, we use the modulus of the wavelet envelope $|\psi(t,s)|$ as a positive-definite weighting function. At each scale, $|\psi|$ defines the local temporal support over which the mean magnetic field is evaluated, allowing the wavelet coefficients to be rotated into parallel and perpendicular components relative to the local mean field direction.
			In our analysis, the data is rotated into a local, field-aligned reference frame, with the directions ($\parallel,\perp_1,\perp_2$) defined with respect to the local magnetic field evaluated separately at each scale. The choice of the perpendicular axes is arbitrary. Here we use the construction introduced by \citet{bieber1996}, which has been used to investigate anisotropy in single-spacecraft measurements, although in the VLISM the applicability of Taylor’s hypothesis is very limited and compressible fluctuations not negligible. The reference system definition is provided in 
			Appendix~\ref{sec:appB}.
			
			The results are shown in Fig.~\ref{fig:spectraLargescale} in terms of compensated power spectra, $P(t,f)f^{5/3}$  (panels~a and~c), where the trace spectrum is
			\[
			P(t,f)=\sum_{j=1}^{3} P_j(t,f),
			\]
			and the magnetic compressibility, defined as $C = P_{\parallel}/P$ (panels~b and~d) for V1 (panels~a and~b) and V2 (panels~c and~d), respectively. The shown intervals also include a year of IHS data, to highlight the difference in turbulence properties from inside to outside the heliosphere. A preliminary version of this figure was first presented at the 22nd AIAC meeting in 2025. The choice to compensate the PSD by the $5/3$ power is arbitrary and does not imply any specific turbulence phenomenology. It is used to facilitate the visualization and to allow us to assess whether the spectral slope is steeper or shallower than the Kolmogorov reference index. 
			
			We first note a clear difference in the turbulence intensity, both in the IHS and in the VLISM, at both spacecraft. At V1, enhanced fluctuations are typically found behind main compression waves, including  \textit{V1-pf2}. It is also possible to identify quasi-periodic structures (with the period from 10 to exceeding 100 days) as localized enhancements of the wavelet PSD persisting over more than a few periods at a given timescale. After 2022, we note a decrease in turbulence intensity, especially at temporal scales longer than 10 days. No dramatic decrease in intensity is observed at the 1 day scale. The magnetic compressibility at scales from 1 to 10 days is relatively low from 2017 to early 2022, with the exception of the post-\textit{pf2} interval, which shows a large compressibility. Afterward, an increase is observed during 2022.5–2024.5. 
			
			There is a remarkable difference in the fluctuation intensity observed by V1 and V2. More specifically, V2 recorded a bulge in turbulence energy at the 1–10 day timescale (compare panels~a and~b in Fig.~\ref{fig:spectraLargescale}), corresponding to a bump in the power spectrum and low magnetic helicity. This feature persists throughout the V2 interstellar data set and is reflected in the relatively high intermittency reported by \citet{burlaga2022}. However, fluctuations are particularly enhanced between 2020.0 and 2021.0, which constitutes a puzzle given the abruptness of the event. The fluctuation level during this interval is well above the typical $1\sigma$ noise level, and the associated scales are far from the frequency range generally affected by magnetometer noise. Moreover, magnetic compressibility is significant throughout the four-year interval analyzed, with the exception of this specific period. However, at this stage we cannot exclude the presence of instrumental artifacts, and a more detailed analysis of turbulence at V2 is left for future work.


			\begin{figure*}[t]
				\centering
				\includegraphics[width=0.9\textwidth]{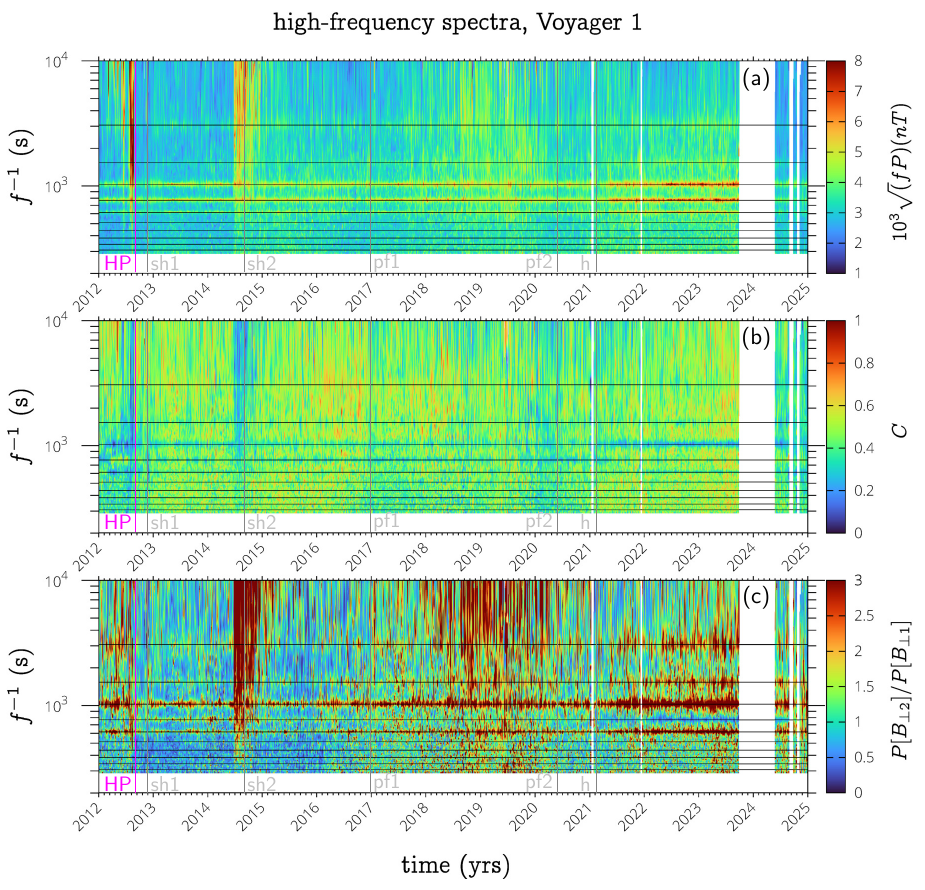}
				\caption{Power spectral density (panel a), magnetic compressibility (panel b), and the ratio of perpendicular magnetic field spectra (panel c) computed from V1 high-resolution data. \label{fig:spectraV1finescale}}
			\end{figure*}
			
			For the fine-scale fluctuation analysis, we inspected 8030 individual segments of contiguous data to construct the time-frequency spectrograms shown in Fig.~\ref{fig:spectraV1finescale}, following the procedure detailed in Appendix~\ref{sec:appB}. The local reference frame here is computed separately for each segment, of typical duration of a few hours. Panel~(a) shows the quantity $\sqrt{Pf}$, representing the fluctuation per logarithmic frequency interval, where $P$ is the trace PSD. Panel (b) shows the magnetic compressibility, and panel~(c) shows the perpendicular power ratio, $P_{\perp 2}/P_{\perp 1}$. 
			
			The scales shown in Fig.~\ref{fig:spectraV1finescale} mostly lie within the noisy range of the MAG data, therefore these results should be interpreted cautiously. Nevertheless, a significant variance anisotropy is observed in specific, relatively long time intervals.
			Discrete spectral lines are observed at $f_k = {k}/({64 \times 48~\mathrm{s}})$.
			The power peaks are sharp and spectrally narrow, remaining fixed over time and exactly aligned with a 64-sample structure of the 48~s averaged data, suggesting an instrumental/spacecraft origin associated with data processing. The identification of these instrumental artifacts is one of the reasons for us to prefer a Fourier-based spectral analysis over a wavelet approach, which typically smears these artifacts. This is particularly important because the lowest-frequency instrumental enhancement lies within the range of frequencies that may be associated with PUI waves.
			
			It is also interesting to note a clear time variability in the fluctuations intensity, as well as a significant peak in magnetic compressibility around $2\times10^{-3}$~Hz throughout most of the interval. Exceptions are the 2014 pre-\textit{sh2} interval and the 2019.5--2020.5 interval, which are characterized by the highest fluctuation intensity, with a strong dominance of perpendicular fluctuations in the $\perp_2$ direction. Moreover, these intervals show the highest fine-scale intermittency as will be shown below.
			
			If instrumental artifacts can be excluded, we propose two possible physical explanations.
			
			\begin{itemize}
				\item The first possibility is that the observed fluctuations are generated in the extensive foreshock of \textit{V1-pf2}, as discussed in Section~\ref{sec:foreshock}.
				\item The second possibility is unrelated to the shock and involves fluctuations generated by instabilities of PUIs born from energetic neutral atoms originating by charge exchange in the fast SW. Inspection of Fig.~\ref{fig:ST_PUI_V1} suggests that the latitude of the sector region in 2018  drops below that of V1, allowing fast neutrals to reach the VLISM at the V1 location. It is not known whether the resulting PUI ring-beam velocity distribution functions are more unstable, as these conditions were not simulated, e.g., in \citet{roytershteyn2019,mousavi2025}, where only PUIs originating in the VLISM from uniform, slow SW conditions were considered.
			\end{itemize}
			
			The ratio of perpendicular spectra shown in Fig. \ref{fig:spectraV1finescale} (b) may also be associated with the wavenumber anisotropy, or 2D fluctuations, in the regions where the ratio exceeds 1.5. However, very large values exceeding 3 may indicate significant deviations from axisymmetry of fluctuations, inapplicability of Taylor's hypothesis, or instrumental artifacts.
			
			
			\begin{figure*}[h!]
				\centering
				\includegraphics[width=\textwidth]{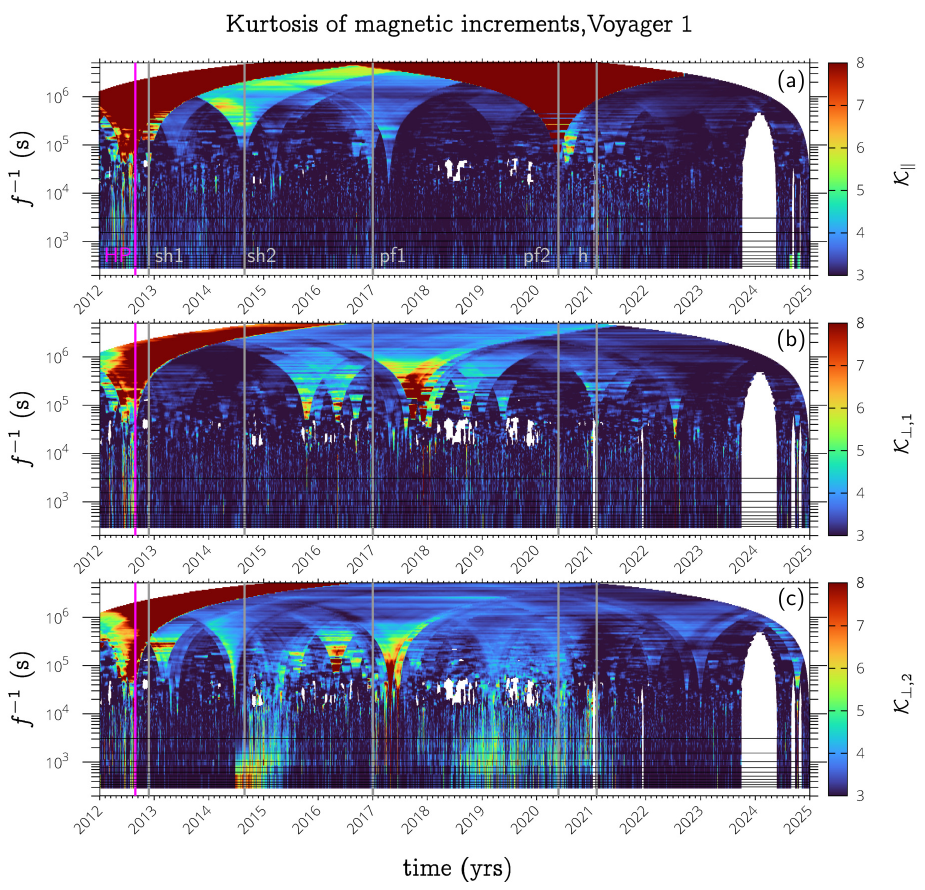} 
				\caption{Spectrograms of the intermittency of magnetic turbulence observed by V1, measured via the kurtosis of the magnetic field increments parallel to the mean field direction ($\mathcal{K}_\parallel$), and perpendicular to it ($\mathcal{K}_{\perp1}$, $\mathcal{K}_{\perp2}$).\label{fig:kurtosisV1}}
			\end{figure*}

			In Fig.~\ref{fig:kurtosisV1}, we quantify the scale-dependent intermittency of V1 magnetic field fluctuations using the increments of $\mathbf{B}$. We follow the approach adopted in our previous studies, but extend the analysis to the full available data set by constructing a running kurtosis. For each magnetic field component $B_j$, we use V1 high-resolution data resampled at a 288~s cadence and cleaned of outliers. The magnetic increment at scale $\tau$ is defined as $\Delta B_j(t,\tau)=B_j(t)-B_j(t+\tau)$. The local kurtosis is then calculated as
			\begin{equation}
				\mathcal{K}_j(t,\tau)=
				\frac{\left\langle \Delta B_j(t',\tau)^4 \right\rangle_{t_w}}
				{\left\langle \Delta B_j(t',\tau)^2 \right\rangle_{t_w}^2},
			\end{equation}
			where $\langle \cdot \rangle_{t_w}$ denotes a moving average over a finite time window  $t_w$ centered at time $t$.
			
			The averaging window is chosen to be proportional to the increment scale by a factor of 50 (i.e., $t_w = 50\tau$), and has no less than $\sim$1000 data points, to ensure proper statistics and a meaningful representation of each scale. This choice produces the typical funnel-shaped structures in the intermittency spectrogram shown in Fig.~\ref{fig:kurtosisV1}, which identify localized small-scale structures, such as current sheets or abrupt jumps in $B$, that lead to intermittency at larger scales. Regions of the spectrogram affected by major data gaps, or by insufficient statistics due to gaps, have been blanked. This analysis covers temporal scales from 288~s up to 60~days.
			
			We emphasize that a global view is necessary for correctly interpreting the Voyager observations. Looking at the kurtosis values calculated by \citet{fraternale2021a}, and the present results shown in Fig.~\ref{fig:kurtosisV1}, it is clear that intermittency as measured by Voyager in the VLISM has never been particularly strong when compared with the typical values observed in either supersonic SW or IHS turbulence before the HP crossing \citep{fraternale2019a}. 
			This does not allow us to conclude that the true intermittency in the VLISM is weak or absent. Fourth-order statistics are strongly affected by the signal-to-noise ratio, the amount and distribution of data gaps, and the binning or filtering applied to the 48-s data. In \citet{fraternale2021a}, we conducted a detailed analysis of the scale-dependent convergence of statistical moments, concluding that at least 2,000 data points are required for reliable estimates of fourth-order moments (see Fig.~11 in that study). Moreover, we would like to make a comment about using  the averaged data for calculations of scale-dependent intermittency. Several studies use 1-hour averaged data to draw conclusions about intermittency at time lags of order 1 hour. This practice is generally problematic, because time averaging can suppress structures at the very scales that define intermittency, which can significantly alter fourth-order statistics. For SW low-noise data, this would typically lead to an artificial reduction of intermittency levels. In the case of Voyager data, the situation is actually more complex due to the high level of noise, and using averaged measurements may help reduce the  noise effects. Conversely, data gaps, when ignored or improperly treated,  may artificially enhance  the intermittency at some scales.
			
			The \textit{observed} intermittency in the VLISM is weak, with kurtosis values rarely exceeding 8. At scales from a few hours up to hundreds of days, intermittency is governed by the presence of steepened, wave-like profiles with quasi-periodic features that are ubiquitous in the VLISM. At these scales, intermittency appears to decrease significantly after the magnetic hump ($\sim$2021.5), which is consistent with both declining solar activity and physical wave damping expected with increasing distance from the HP. However, intermittency does not vanish at frequencies below $10^{-5}$~Hz, even in the most recent data. Moreover, the parallel component generally shows the smallest values of $\mathcal{K}$. Finally, these scales are significantly larger than the estimated scales of the ion-subrange regime of turbulence.
			
			More interesting is the behavior of intermittency at small scales of 1~hr or less. Figure~\ref{fig:kurtosisV1} shows that intermittency is dominated by one perpendicular component ($\perp_2$) and that, excluding the interval prior to the HP crossing, it is significant only during two interstellar time intervals: the first from $\sim$2014.5 to $\sim$2015.5, associated with \textit{V1-sh1} \citep{fraternale2020a}, and the second during 2018-2021.5, with peak intensity in 2019. This corresponds to the intermittent interval identified in our previous study,  and also by \citet{burlaga2024a} who used the hourly increments. A marked drop in 1~hr-scale intermittency occurs after 2022, after which the increments statistics become near-Gaussian.
			
			Therefore, our results highlight the strongly time-dependent nature of both intermittency and turbulent fluctuation intensity. There is no justification for using the intermittency measures alone to draw conclusions about the nature of \textit{V1-pf2} without considering the global context. The region immediately behind the shock is in fact intermittent, and the 2022–2025 interval is not the only one characterized by near-Gaussian statistics at the 1~hr scale; for example, the 2013–2014 interval, which was also close to the HP, showed a similar behavior. 
			\begin{figure*}[h!]
				\centering
				\includegraphics[width=\textwidth]{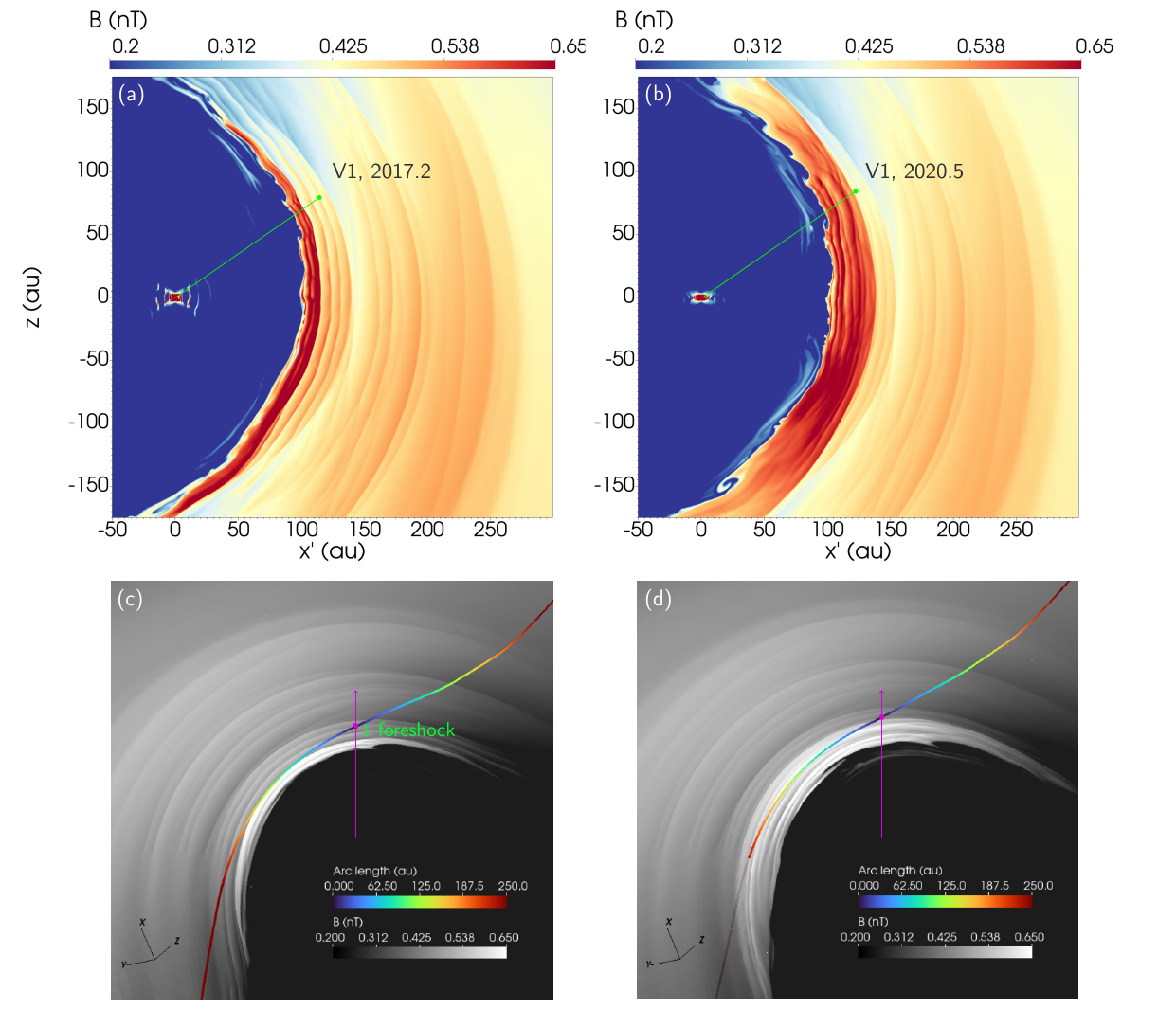} 
				\caption{The foreshock region of \emph{pf2}. Panels (a) and (b) show the meridional distributions of the magnetic field strength in the V1-$z$ plane, in 2017.3 and 2020.5, respectively, from simulation \texttt{B1}, illustrating the pressure front  propagating through the VLISM. The bottom panels show the same distributions in a plane containing the local magnetic field line and V1. The magnetic field line is colored according to the distance from the spacecraft. The figure shows that V1 could be magnetically connected to the shock as early as 2017 and illustrates the broad spatial extent of the foreshock region at the time when local magnetic field line at the spacecraft becomes tangent to the shock surface.\label{fig:foreshock_streamline}}
			\end{figure*}
			
			\subsection{The foreshock of \textit{V1-PF2}}\label{sec:foreshock}
			
			The top panels of Fig.~\ref{fig:foreshock_streamline} show two snapshots of the magnetic field strength at 2017.2 and 2020.5, respectively, in the meridional  plane containing the \textit{z} axis and V1. These panels illustrate the propagation of \textit{V1-pf2} and the effects of heliopause rippling on the shock geometry in this plane. The bottom panels show the same distributions in the plane containing V1 and the local magnetic field line. The normal to this plane was calculated using a best-fit procedure applied to the magnetic field line within radial distances of 200~au and, when expressed in Ecliptic J2000 coordinates, is $\mathbf{N}=(0.67013,0.08169,0.73773)$ and $\mathbf{N}=(0.67245,0.08858,0.73482)$ for 2017.2 and 2020.5, respectively.

			Here we investigate the geometric extent of the foreshock in the SC-24 perturbation by identifying the time when V1 becomes magnetically connected to the shock, i.e., when the magnetic field line passing through the spacecraft becomes tangent to the shock surface. Interestingly, this appears to occur around 2017.2, as shown in panel (c). In panel (d), we show the moment when the shock reaches the spacecraft. 
			The magnetic field line is colored according to the distance from the spacecraft along the line. It can be seen that, at the tangency condition of panel (c), 
			V1 is as far as $\sim130$ au from the magnetic connection point. The resulting foreshock region is very broad, 
			spanning up to $\sim15$ au. Clearly, the presence of a geometric foreshock does not guarantee observable foreshock signatures. We therefore attempt to identify possible observational signatures associated with this foreshock. If the VLISM shock was able to accelerate ions, the enhanced fine-scale turbulence identified between 2018 and 2020 and shown in Fig.~\ref{fig:spectraV1finescale} might represent one such signature, similar to what was observed ahead of \textit{V1-sh2}. Moreover, we recall that three electron plasma oscillation (EPO) events were recorded between 2017.5 and 2019 (EPO 6, 7, and 8) in \citet{kurth2024}, with EPO~6 being the most intense. In 2017.5, V1 recorded also an increase in energetic particle fluxes, including MeV GCRs measured by V1/CRS and 40–139~keV protons measured by V1/LECP. Because no shock was directly observed by MAG during the 2017.5–2020 interval, we suggest that this event might be related to the foreshock region of \textit{V1-pf2}, as its timing is reasonably close to the estimates provided by our models, given the large uncertainties involved.

			\section{Conclusions}\label{sec:conclusions}
			
			We have used an updated, data-driven version of our five-fluid global model of the SW--LISM interaction to  interpret Voyager~1 and~2 observations in the outer heliosheath and to link the interstellar transients to their SW origins. For the first time, we simulated more than four solar cycles using inner BCs imposed at 1~au and which combine high-cadence OMNI data with the IPS-based SW model of \citet{porowski2023}  
			and \citet{porowski2024}.  This approach allows the heliosphere to evolve self-consistently over multiple cycles and introduces an additional variability that must be accounted for when interpreting spacecraft observations.
			
			A central result is that the solar cycle drives a global VLISM response in the form of large-scale compressions that persist for several years, can reverse the radial component of the interstellar plasma flow, and propagate to hundreds of au. In the simulations, each cycle produces one major perturbation that involves both the motion of the fast/slow SW boundary and the post-maximum increase in the SW dynamic pressure at 1~au. This mechanism provides a convincing interpretation of the \textit{V1-pf2} pressure front and explains the relatively strong magnetic field values recorded by Voyager~1 since mid-2020, which we interpret as a global SC-24 disturbance. In our runs, \textit{V1-pf2} results from the merging of two shocks at the time of encounter, while a delayed pressure front forms the magnetic ``hump'' described by \citet{burlaga2023}. We quantitatively reproduce the non-decreasing behavior of $|\mathbf{B}|$ behind \textit{pf2}, which arises from the pileup of multiple compressible wave fronts, each generated at the heliopause during 2016-2018 every 4-6 months, producing a post-shock ramp that is shallower than the standard N-wave profile typical of the relaxation of GMIR-related interstellar shocks.  In the models, this ``hump'' wave  is shaped by continued interaction with higher-frequency fluctuations and could be traced back to a limited number of major dynamic-pressure enhancements in the supersonic SW observed along the NH direction between mid-2016 and mid-2017.

			Quantitative comparisons with the spacecraft time series require simulations of a  spacecraft moving through the simulation domain. Systematic offsets in heliopause location should also be taken into account. Our models underestimate the heliosheath width by a few au, which we will address in the future simulations. In addition, the HP corrugation due to instabilities and reconnection introduces intrinsic variability, so that identical inner BCs can lead to different wave arrival times and amplitudes in the VLISM, highlighting the need for uncertainty quantification and ensemble-type simulations.
			
			We predict that the magnetic field strength at V1 will not recover to pre-\textit{pf2} levels until at least $\sim$2030. Moreover, models with $B_\lism=3.50~\mu$G generally provide a better match to the observed magnetic field strength due to the weaker background radial gradient. Continued Voyager~1 measurements may therefore help further constrain pristine LISM conditions when interpreted with global, time-dependent models.
			
			Along the V2 trajectory, the simulations indicate that an enhanced southern-latitude compression, transient phasing relative to the SC-24 global disturbance, and the timing of a nearby shock at the HP crossing collectively explain the strong magnetic field  near the HP and its subsequent decline. The absence of a strong \textit{V1-pf2} analog in the V2 observations is a consequence of the SC-24 perturbation having been generated before V2 entered the VLISM.

			We conducted a turbulence and intermittency analysis of the full publicly available Voyager data sets, confirming that VLISM magnetic fluctuations are strongly time dependent. The apparent reduction of intermittency and turbulence intensity after the V1 ``hump'' is consistent with declining solar activity and increasing distance of the spacecraft from the HP, which in our models starts retracting around mid-2021. At relatively fine scales (spacecraft-frame timescales below $\sim$1~hr at V1), intermittency is detected only during two interstellar intervals, the second of which occurs during 2018--2021.5 and whose initial phase was first identified by \citet{fraternale2021a}. A near-Gaussian statistics observed during 2022--2025 is not unique and does not imply the absence of intermittency; rather, they are consistent with a low signal-to-noise ratio.

			Finally, the global geometry in all models suggests that V1 may have been magnetically connected to the \textit{pf2} shock as early as $\sim$2017.2, implying the presence of an extended foreshock region that could explain the enhanced turbulence signatures and possibly other observations such as radio emissions recorded by V1/PWS.
			
			Overall, our results show that solar-cycle-driven global compressions provide a unified explanation for \textit{V1-pf2}, the sustained post-front magnetic-field enhancement, the magnetic hump, and the differences between V1 and V2 observations. The simulations also indicate that major perturbations decelerate with distance, making it unlikely that Voyager~1 will be overtaken by the main SC-25 disturbance, should such an event occur. In contrast, V2 is expected to experience stronger solar-driven activity in the coming years, including at least two compressions before the end of 2026 and a major SC-25 event of the same nature as \textit{V1-pf2} around $\sim$2030. 
			
			 A number of important issues remain unsolved. One of them concerns the dissipation mechanism in weakly collisional shocks observed in the VLISM and, consequently, determine their structure, ability to reflect ions and electrons, and generate radio emission.  The dissipation mechanism plays a substantial role in shock-related heating of the OHS plasma. 
			To address this issue, one would  need a better understanding of the ion and electron distribution functions in the OHS, so the problem cannot be analyzed within the MHD framework alone. The second major issue concerns the evolution of the magnetic field direction in the VLISM. In fact, the VLISM may be more dynamic than the current global models can capture. Extrapolations of the data reveal inconsistencies: (i) between the Voyager data and the direction inferred from IBEX; (ii) between the observations and existing global models; and (iii) possibly, between the two spacecraft, although V2 is still too close to the HP to allow for a meaningful extrapolation of the field direction. If the current rotation of the field is attributable to the pristine interstellar turbulence, this would imply either a gap in our understanding of the  fluctuation spectrum in this region or in the way the LISM turbulence is processed by the bow wave. Alternatively, if the heliopause and heliospheric dynamics are responsible for generating such large-scale fluctuations, to capture them, one would need to enhance the current global models  with some additional physics. Future progress will require improved inner BCs, MHD-plasma/kinetic-neutral modeling, and systematic uncertainty quantification to better constrain the pristine LISM conditions. The results presented here and their future extensions, will be important for the interpretation of the forthcoming IMAP data.

			\appendix
			\section{Methods}\label{sec:appA}
			
			\subsection{Global modeling framework and simulation setup}  \label{sec:appA1} 
			Numerical simulations were performed on a Cartesian grid using the multifluid version of the Multi-Scale Fluid-Kinetic Simulation Suite \citep[MF-FLUKSS, see][and references therein]{pogorelov2008d,pogorelov2014} with adaptive mesh refinement (AMR). Specifically, we used a modified version of the model published in \citet{bera2023a,bera2025}, where PUIs were treated as a separate, comoving fluid. \citet{bera2025} discusses the consequences of adopting this approximation ,as opposed to not adopting it, in more detail.
			The modifications implemented here were explained in Sec.~\ref{sec:framework}. The use of four populations of neutral hydrogen allows us to distinguish between the pristine LISM atoms from their population created in  the VLISM. This makes the H profiles in the OHS closer to those obtained from MHD/kinetic-neutrals models \citep{alexashov2005,pogorelov2009c}. While we do not solve separate equations for \HeI and \alp ions in this study, nor electrons, \citep[as is done in MHD/kinetic models, see, e.g.,][]{fraternale2023,fraternale2024b,fraternale2025a}, the presence of helium ions is approximately accounted for by adjusting the mass and pressure in the proton-only model. The feedback of helium ions due to charge exchange cannot be included, but this is of secondary importance for this study.
			These modifications are important as they affect the hydrogen filtration, electron density in the LISM, and correct positions of the heliospheric boundaries.
			Even though conceptually simple, such a modification requires a careful implementation, because it is not sufficient to simply use a reduced proton mass in the analytical charge-exchange formulae. Moreover, some assumptions are inevitably required, as detailed below.
			
			We define the following ratios at the inner and outer boundary:
			\begin{gather}
				r^\mathrm{SW}_{n,\alp}\equiv\frac{n_{\alp,1 \au}}{n_{\p,1 \au}}, \\
				r^\mathrm{SW}_{n,\HeI}\equiv 0,\\
				r^\mathrm{LISM}_{n,\HeI}\equiv\frac{n_{\HeI, \lism}}{n_{\p, \lism}}, \\
				r^\mathrm{LISM}_{n,\alp}\equiv 0.
			\end{gather}
			
			During simulations, it is then assumed that the helium ion to proton density and temperature ratios remain constant in the entire region filled with the SW and LISM plasma, and is equal to the respective boundary ratios. In the LISM regions, this approximation is rather accurate due to the weakness of helium charge exchange and the presence of Coulomb collisions. In the SW, for the purpose of this study it is mostly important to consider the mass of alpha particles and we neglect the presence of \HeI ions.
			
			Therefore, assuming that everywhere $T_\e=T_\pth$  \citep[as in][]{pogorelov2016,bera2023a} and $T_\HeI=T_p$ and $T_\alp=T_\pth$, and resolving the equations for the plasma mixture and \puis as in \citep{bera2023a}, the core proton properties in the SW regions are obtained as
			
			\begin{gather}
				n_\pth({\bf x},t) = \frac{\rho}{m_\p}\left[ 1 + \frac{m_\He}{m_\p} \left( r^\mathrm{R}_{n,\HeI} + r^\mathrm{R}_{n,\alp} \right) \right]^{-1} - n_\pui, \\
				p_\pth({\bf x},t) = \frac{p-p_\pui}{(2+\alpha_\pui)+(1+\alpha_\pui)(2 r^\mathrm{R}_{n,\HeI} +3 r^\mathrm{R}_{n,\alp})}, \\
				\text{where} \nonumber \\
				r_{n,s}^R =
				\begin{cases} 
					r_{n,s}^{SW} & \text{in the SW regions}; \\ 
					r_{n,s}^{\lism} & \text{in the LISM regions},
				\end{cases}
			\end{gather}
			where $\alpha_\pui=n_\pui/n_\pth$,  $p=p_\pth+p_\pui+p_\alp+p_\HeI+p_\e$  is the thermal  pressure of plasma, and $\rho=\rho_\pth+\rho_\pui+\rho_\HeI+\rho_\alp$ is the plasma density.  Note that the above expressions are  used simply to recover the thermal proton properties from the evolved variables whenever needed, e.g., in the calculation of charge-exchange reactions or in postprocessing tasks. Therefore, these are algebraic relations introduced for practical purposes.

			We do not use any turbulence transport model in the current simulations. Therefore, the core SW proton heating due to PUIs in the supersonic SW is relatively simple and obtained by decreasing, by a fixed fraction, the source term in PUI pressure equation, i.e., the pressure rate due to charge exchange, as done in \citet{bera2025}. In the data driven simulations, it was sufficient to use a fixed fraction equal to 3.75\% instead of the 5\% needed in steady-state or nominal solar cycle models.
			We run global simulations using Cartesian grids, in a cubic domain of side 1680 au, and Sun-centered coordinates $x\in[-1000, 680], y,z \in[-840,840]$. For the final data-driven runs, we used a base grid of size $\Delta x=\Delta y=\Delta z = 10 \au$, and 7 additional levels of refinement. The inner BCs are imposed at 1 \au and the finest grid is 0.075 au cubed.

			We have designed the grid to have a uniform resolution of 0.31 \au in the whole upwind side of the heliosphere extending up to 200 au into the LISM.

			While the spherical grids used in the simulations of \cite{kim2017b}  \citep[see also][]{pogorelov2021,zirnstein2024} have a higher resolution in the radial direction, they also have a high aspect ratio, resulting in a lower resolution in the azimuthal and latitudinal directions.  Our choice of using Cartesian grids was motivated by the need to avoid suppressing of, or biasing in, the development of large-scale structures and anisotropy of resolved fluctuations, to capture their impact on VLISM perturbations.

			\begin{figure*}[t!]
				\centering
				\includegraphics[width=0.8\textwidth]{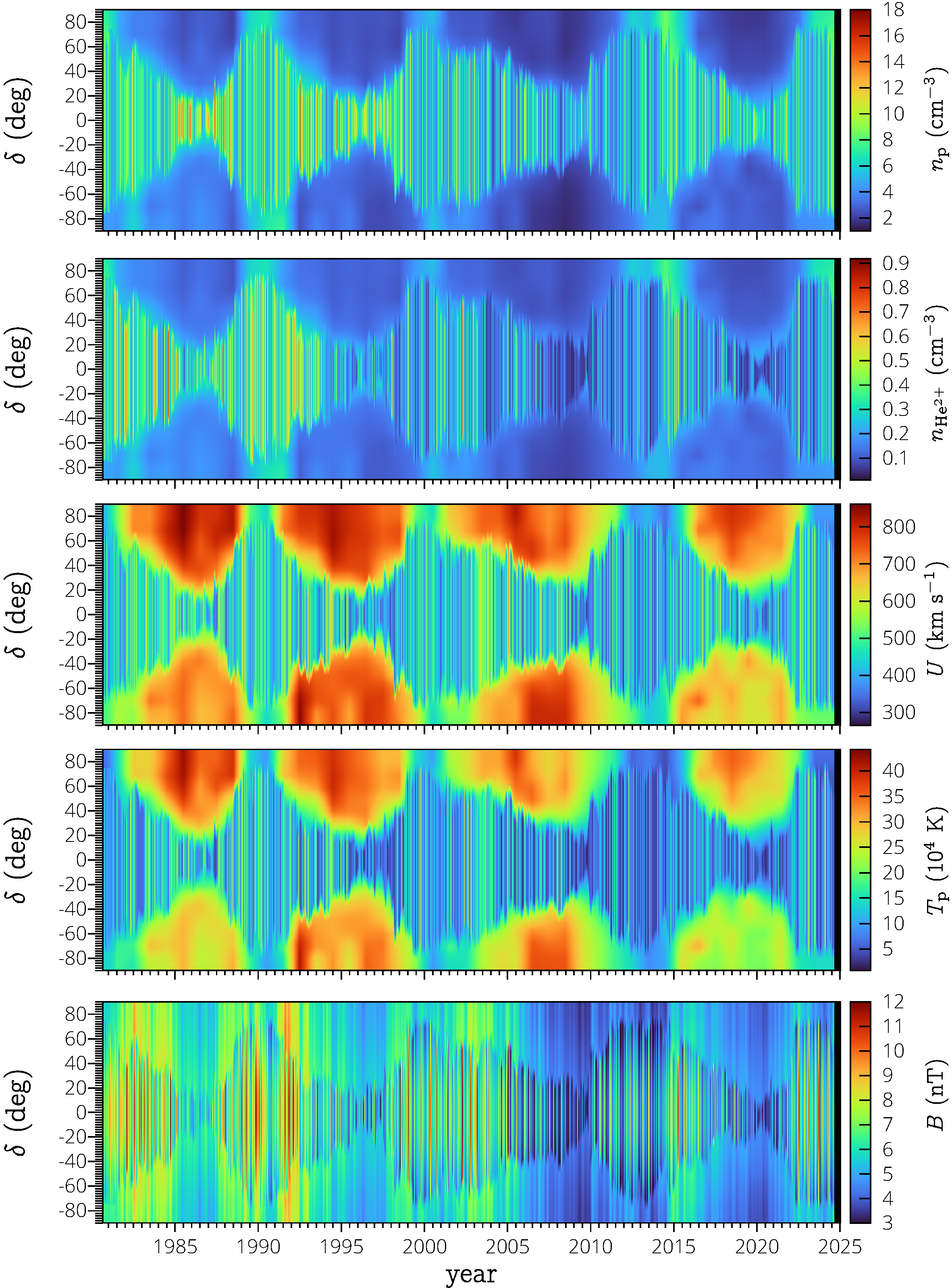}
				\caption{The boundary conditions used at 1 au for all simulations presented in this study, as described in Appendix \ref{sec:appA2}.\label{fig:BCs}
				} 
			\end{figure*}
			
			\subsection{Inner boundary conditions for the global models}\label{sec:appA2} 
			
			Our inner BCs at 1 \au, shown in Fig.~\ref{fig:BCs} are based on Interplanetary Scintillation, OMNI and Ulysses, and WSO data, and implemented as follows. We used the most recent model of the heliolatitude behavior of the SW density and speed as functions of time \citep{porowski2023}. The SW speed is based on IPS when they are available.  When not available, a modified methodology from \citet{porowski2024} is used. This data consists of time-latitude maps with the time resolution of a Carrington rotation. 
			
			The possibility to run several solar cycles is the key to our modeling results. In addition to the speed and density data, we need to specify the magnetic field vector, the latitudinal extent of the heliospheric current sheet, helium ion density, and plasma temperature. This is done as follows.
			
			\begin{itemize}
				\item Latitudinal extent of the HCS.
				We use the Wilcox Solar Observatory  data (more precisely, their ``classic'' model, which is North-South asymmetric) \footnote{http://wso.stanford.edu/} (Carrington rotation resolution).
				
				\item HMF vector. We use the HMF magnitude from the OMNI data at 1-day resolution and construct the HMF vector using the Parker model. This is used in the region occupied the the HCS only, while at higher latitudes a fixed magnitude of 5.5 nT is used (the vector components still following Parker), to fit reasonably the  Ulysses data set. 
				
				\item Proton density. OMNI data at 1-day resolution is used in the HCS region (ecliptic data is extended at higher latitudes, similarly to what was done by \citet{kim2017b} and \citet{zirnstein2024}), while the \citet{porowski2023} IPS-based model is used at higher latitudes. A smooth transition is ensured by a 10 degree smoothing. 
				
				\item Alpha particle density. OMNI data at 1-day resolution is used in the HCS region, and a fixed alpha to proton density fraction of 4.5\% is used at higher latitudes, to match the Ulysses data with a reasonable accuracy.
				
				\item Temperature. We use OMNI data at 1-day resolution in the HCS region, and an analytic expression based on the correlation between temperature and speed derived using OMNI data. In contrast to  \citet{pogorelov2013b}, who used different analytic expressions for each of the solar cycles observed by Ulysses, here we derive a simple formula based on the temperature-speed correlation, which we use in the fast wind regions as well. The expression reads 
				\[ \begin{split}
					T_{1\au} &= \max\Bigg(10^4,  \Big[a_0 + a_1 \left(\frac{V}{V_0}\right)^{b_1} \Big] \Bigg)~(K)
				\end{split}
				\]
				with the constants  $a_0=-4.162\times10^4$ K, $V_0=400$ km~s$^{-1}$, $a_1=1.1158\times10^5$ K, $b_1=1.864$.
			\end{itemize}
			
			Because a key objective of our work is to simulate several solar  cycles to allow the solution to develop large-scale, cycle-related global features, we extend the available data by assuming periodicity, ensuring that we use data corresponding to an integer number of solar cycles. Therefore, rather than using the entire dataset available since 1976, we select 44 years of data, spanning from 1980.65 to 2024.65. Once the simulation time reaches the end of this range, the four cycles of data are repeated indefinitely. To avoid discontinuities in the physical quantities, we make the 1980.65 data match the 2024.65 data and apply a linear smoothing over the first half year (1980.65–1981.0). It is important to note that we do not modify the more recent data.

			The time-latitude BC data need to be transformed into spatial BCs around our inner boundary at 1 au, filling the azimuthal directions where data is not available. With such data, the most correct way to do this is by  using the ``sliding window'' method, as done as by \citet{kim2016}. At each moment of time $t$ during the simulation, the data from the table is wrapped around the 1 au sphere in such a way that 
			\begin{gather}
				\mathbf{\mathcal{B}}(\mathbf{x}, t) = \mathbf{\mathcal{T}}(t', \delta) \quad \text{for} \quad \mathbf{x} \in \partial \Gamma\\
				t' = t + \Omega^{-1} \left[\phi_E(t) - \phi(\mathbf{x})\right]
			\end{gather}
			where $ \Omega $ is the Sun's angular rotation frequency, $ \phi $ is the azimuthal coordinate, $ \phi_E(t) $ is the azimuth of Earth, $ \mathbf{\mathcal{B}} $ represents the boundary conditions, and $ \mathbf{\mathcal{T}} $ is the lookup table containing the boundary data. This table is a 3D array that is read once at the beginning of the simulation and stored in the code.  In summary, the actual data is oriented towards in the real (time varying) position of Earth (where OMNI and IPS observations are taken), while the future and past data is used to fill the other azimuthal directions consistently with the Sun's rotation. This necessarily creates a discontinuity at 180 deg away from the $\phi_E$ direction, and a smoothing over a few degrees in longitude is applied to remove it.  Some studies employ a ``vibrating sphere'' approach \citep{washimi2011}, where the boundary conditions vary with time either in a spherically-symmetric of axially-symmetric way. Such approaches are prone to artifacts and cannot produce solar realistic wind structures that result in the hump-like solutions.
			
			These inner boundary conditions are applied to the discretized surface of the spherical volume $ \Gamma $ with a 1~au radius. Unlike the spherical grid case, we do not use ghost cells here, and the Cartesian grid exists inside $ \Gamma $, where the plasma solution is replaced at each time step with analytic expressions. Instead, the neutral fluid equations are solved throughout the entire volume of $ \Gamma $, where charge-exchange and photoionization act as sinks for H atoms.

			\section{Data analysis: Spectrogram processing methodology}\label{sec:appB} 
			
			The final time--frequency spectrogram of Fig.\ref{fig:spectraV1finescale} in this work is obtained through a two-stage procedure consisting of segmented spectral estimation from the original time series followed by the uniform time regridding and temporal smoothing. This is necessary to handle irregular sampling and extended data gaps while preserving physically meaningful spectral information. 
			
			The first part essentially consists of the method labeled as '\texttt{SS}' described in the Appendix of \citet{fraternale2019a}: the input time series $B_j(t)$, with $n_{\mathrm{comp}}$ components, is first partitioned into contiguous segments separated by data gaps. Two successive samples are considered contiguous when their time separation is smaller than a threshold $t_{\mathrm{gap}} = 2\,\Delta t$, where $\Delta t = 144\,\mathrm{s}$ is the cadence of the original time-averaged data used in this study.
			Each segment is defined by its start and end times $(t_{k_1},t_{k_2})$ and is assigned the time $  
			t_{\mathrm{ave}} = (t_{k_1}+t_{k_2})/2$. The segments containing fewer than a minimum number of data points are discarded. Within each retained segment, the data points are interpolated onto a strictly uniform grid with spacing $\Delta t$ using linear interpolation, thereby filling only small internal gaps.
			
			For each segment, the vector time series are then rotated into a segment-dependent reference frame (i.e., the $\parallel,\perp_1,\perp_2$ frame described below).

			The parallel direction is defined by the unit vector along the local mean magnetic field,
			$\hat{\boldsymbol{e}}_{\parallel} =
			{\langle \boldsymbol{B} \rangle}/{|\langle \boldsymbol{B} \rangle|}.
			$
			Using a fixed, average relative flow direction (accounting for the spacecraft velocity) in RTN coordinates
			\[
			\hat{\boldsymbol{e}}_{V}=(\mathbf{U}_\mathrm{plasma}-\mathbf{V}_{V1})/|\mathbf{U}_\mathrm{plasma}-\mathbf{V}_{V1}|=(-0.935\; 0.081\; 0.346),
			\]
			derived using the mean plasma velocity from 2012.5 to 2020 from the simulations (see Fig.~\ref{fig:ST_Ucomps_V1}), the perpendicular directions are defined as
			\[
			\hat{\boldsymbol{e}}_{\perp 2} =
			\frac{\hat{\boldsymbol{e}}_{V} \times \hat{\boldsymbol{e}}_{\parallel}}
			{|\hat{\boldsymbol{e}}_{\parallel} \times \hat{\boldsymbol{e}}_{V}|},
			\qquad
			\hat{\boldsymbol{e}}_{\perp 1} =
			\hat{\boldsymbol{e}}_{\perp 2} \times \hat{\boldsymbol{e}}_{\parallel}.
			\]
			
			Note that the $\perp_1$ direction is the one closest to the sampling direction (and to the $\hat{\mathbf{R}}$ direction of RTN) and corresponds to what was historically referred to as the ``quasi-parallel'' direction, while $\perp_2$ corresponds to the ``perpendicular'' direction in the nomenclature of Bieber et~al (and is close to $\hat{\mathbf{N}}$ ), which refers to the sampling direction and not to the magnetic-field orientation.

			A tapering window must be applied to each segment before computing the discrete Fourier transform, in order to smoothly reduce the signal to zero at the segment boundaries, thereby minimizing spectral leakage \citep[we use the Hann window, but several alternatives exist; see, e.g.,][]{harris1978}. The PSD includes a correction factor depending on the chosen window. The resulting local frequency grid is $f_{\mathrm{loc}} = j/(m\,\Delta t)$ with $j = 1,\ldots,m/2$, 
			where $m$ is the number of samples in each segment. Because different segments may have different lengths and therefore different frequency resolutions, each segment spectrum is interpolated onto a common frequency grid $f$ using log--log interpolation, with undefined values  where interpolation is not possible. In addition to the raw spectrum $P$, a frequency-smoothed spectrum $P_{\mathrm{smo}}$ is obtained by applying an adaptive filter to the local spectrum, with the smoothing window increasing with frequency.

			The collection of spectra $\{P(f,t_{\mathrm{ave}})\}$ constitutes an initial spectrogram sampled at irregular times.
			
			The spectrogram is then interpolated onto a uniform time grid with cadence $\Delta t_{\mathrm{uni}}=6$ hours. For visualization purposes only, short gaps within individual frequency channels are filled by linear interpolation prior to this step, while no extrapolation is performed. Large data gaps are explicitly identified using the original time series and all values on the uniform time grid that correspond to these gaps are set to undefined, so that neither interpolation nor smoothing fills these gaps.
			
			Finally, a temporal smoothing is performed by convolving the uniformly sampled spectrogram with a boxcar window of 1 day length.
			The smoothed spectrogram $\tilde{P}(f,t)$ is computed as
			\begin{equation}
				\tilde{P}(f,t) =
				\frac{\sum\limits_{|t'-t| \le T_{\mathrm{smooth}}/2}
					P(f,t')\,M(t')}
				{\sum\limits_{|t'-t| \le T_{\mathrm{smooth}}/2} M(t')},
			\end{equation}
			where $M(t')$ is a binary mask equal to unity for valid data points and zero otherwise.

			\subsection{Supplemental figures}
			
\begin{figure*}[t!]
				\centering
				\includegraphics[width=0.95\textwidth]{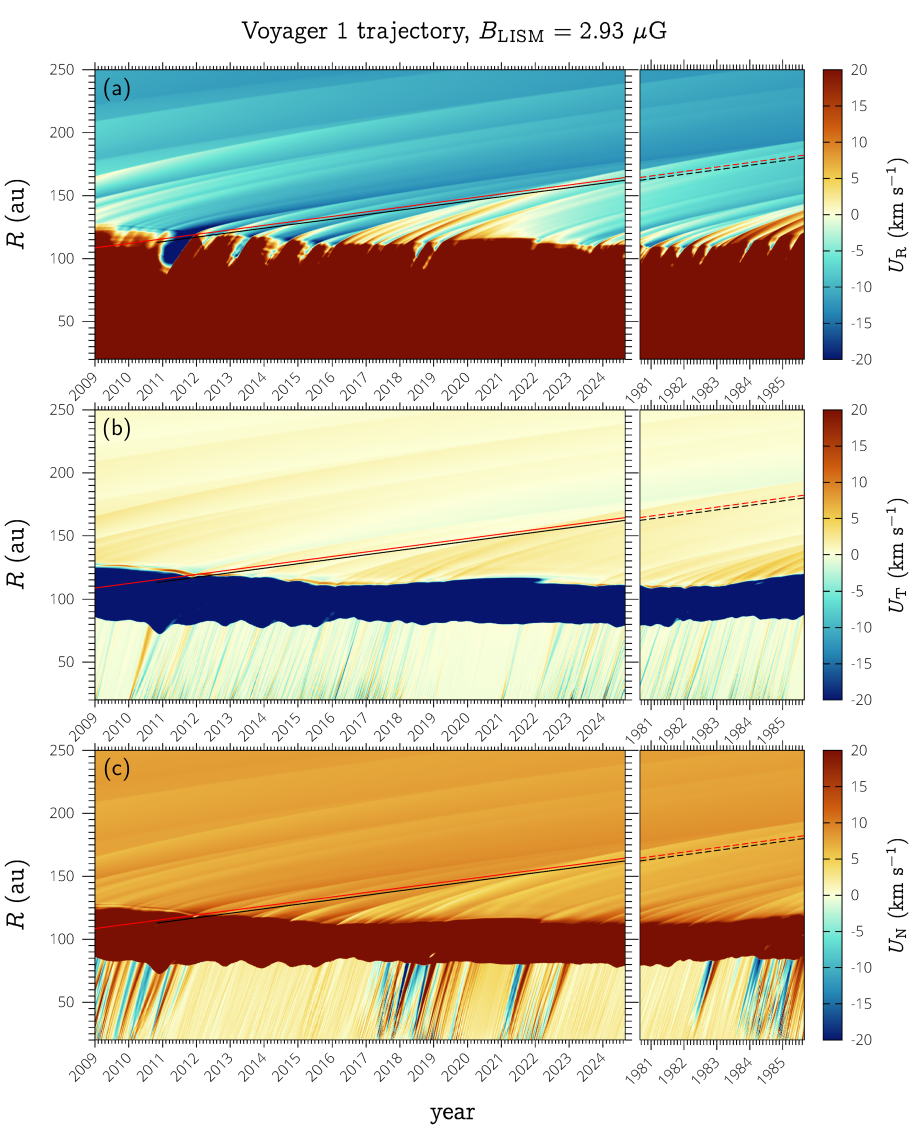}\vspace{-10pt}
				\caption{The RTN components of the velocity vector along the V1 trajectory from simulation \texttt{A}.\label{fig:ST_Ucomps_V1} } 
			\end{figure*}

\begin{figure*}[t!]
				\centering
				\includegraphics[width=0.95\textwidth]{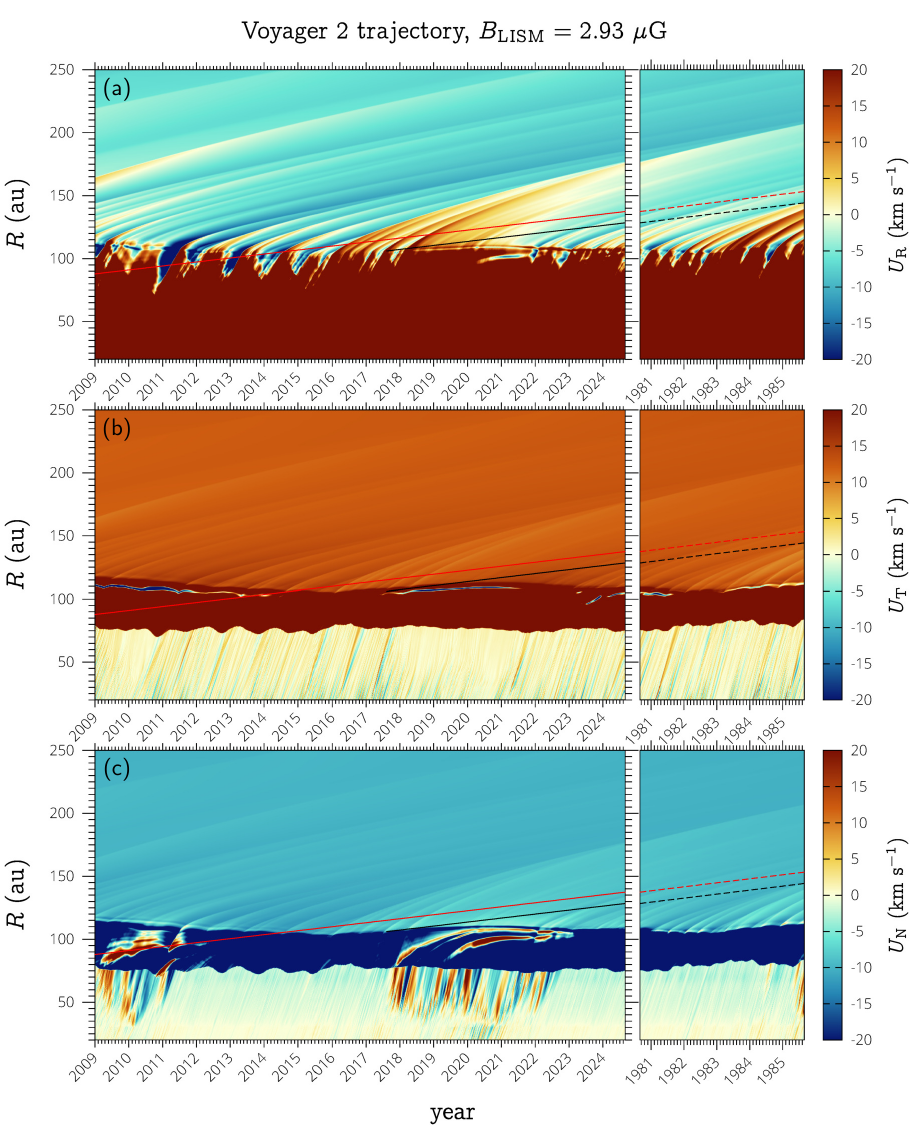}\vspace{-10pt}
				\caption{The RTN components of the velocity vector along the V2 trajectory from simulation \texttt{A}.\label{fig:ST_Ucomps_V2} } 
			\end{figure*}

			We additionally provide a few supplemental figures that support the main results discussed in the text. Figures~\ref{fig:ST_Ucomps_V1} and~\ref{fig:ST_Ucomps_V2} show the plasma velocity components, expressed in the RTN frame, along the V1 and V2 trajectories, respectively. The focus is on the VLISM, as a result, the inner heliosheath region appears saturated in these plots.
			
			Figure~\ref{fig:ST_B_MF015pui} shows the PUI density, temperature, and pressure distributions along the V1 trajectory from simulation~\texttt{B1}, highlighting the time variability of PUIs born in the VLISM.

			\begin{figure*}[t!]
				\centering
				\includegraphics[width=0.95\textwidth]{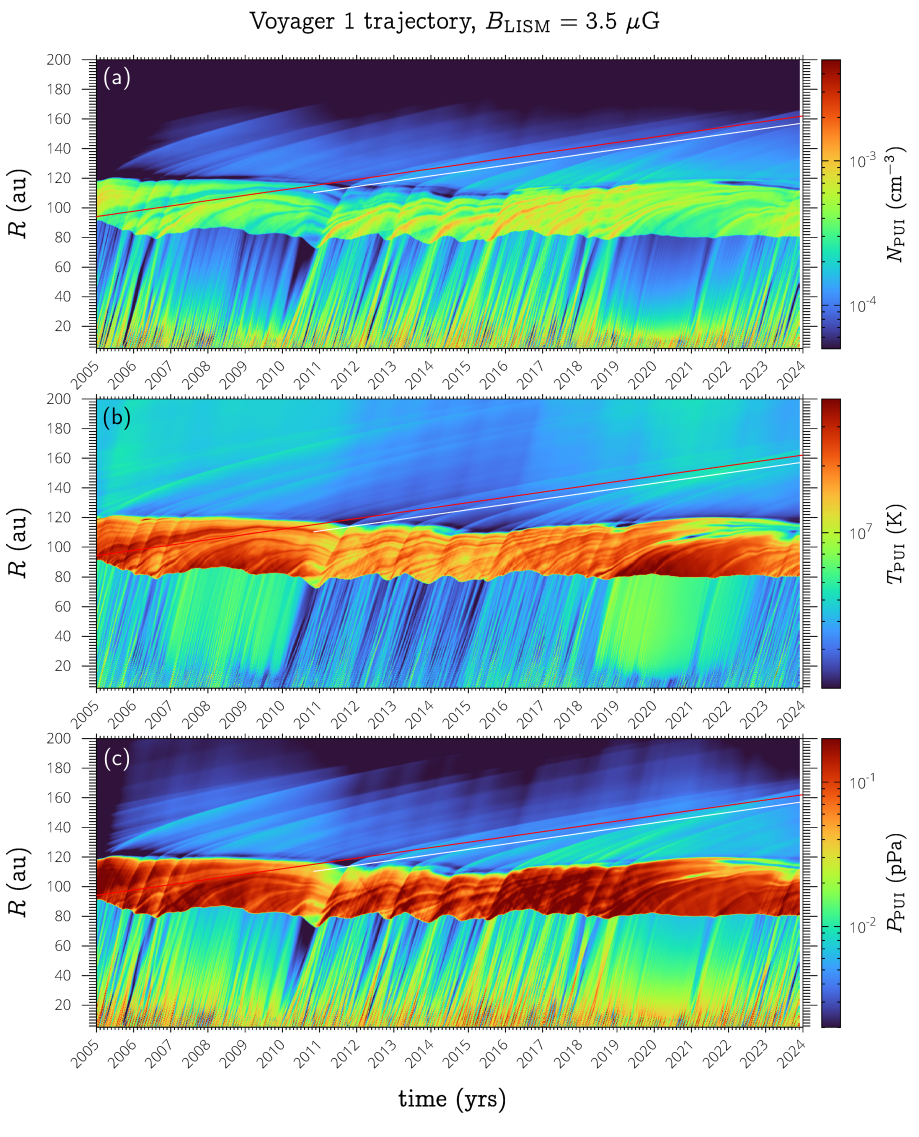}\vspace{-10pt}
				\caption{Space-time distributions of PUI density, temperature, and pressure along the V1 trajectory from simulation \texttt{B1}.\label{fig:ST_PUI_V1}} 
			\end{figure*}

			\begin{acknowledgments}
				This work is supported by NASA grants  80NSSC24K0267 and 80NSSC22K0524, and NSF grant 2512085. M.B. was supported by Polish National Science Centre (NCN) grant 2023/51/B/ST9/01921. We acknowledge the Texas Advanced Computing Center (TACC) at The University of Texas at Austin for providing HPC resources on Frontera supported by NSF award CISE-OAC-2031611, and on Stampede3 supported by ACCESS project MCA07S033.
				Supercomputer time allocations were also provided by NASA High-End Computing Program awards SMD-17-1537 and SMD-23-67004198. We also acknowledge the IBEX mission as part of NASA’s Explorer Program (80NSSC18K0237). FF and NP are grateful to the International Space Science Institute (ISSI) in Bern, for its support through ISSI International Team project \#23-574 ``Shocks, Waves, Turbulence, and Suprathermal Electrons in the Very Local Interstellar Medium''.  
			\end{acknowledgments}
			
			\FloatBarrier

			\bibliographystyle{aasjournalv7}

		\end{document}